\documentclass[a4paper,fleqn,usenatbib]{mnras}

\usepackage{times}

\usepackage[T1]{fontenc}
\usepackage{ae,aecompl}
\usepackage{graphicx}	
\usepackage{amsmath}	
\usepackage{amssymb}	
\newcommand\gsim{~\lower.5ex\hbox{$\buildrel > \over \sim$}~}

\title[Luminous Ly$\alpha$ emitters at $z\sim$\,2--3]{The nature of luminous Ly$\bf \alpha$ emitters at $\bf z\sim2-3$: maximal dust-poor starbursts and highly ionising AGN\thanks{Based on observations obtained with the William Herschel Telescope, program: W16AN004; the Very Large Telescope, programs: 098.A-0819 \& 099.A-0254; and the Keck II telescope, program: C267D.}}

\author[D. Sobral et al.]{David Sobral$^{1,2}$\thanks{E-mail: d.sobral@lancaster.ac.uk}, Jorryt Matthee$^{2}$, Behnam Darvish$^{3}$, Ian Smail$^{4}$, Philip N. Best$^{5}$,  \newauthor Lara Alegre$^{5,6}$, Huub R\"{o}ttgering$^{2}$, Bahram Mobasher$^{7}$, Ana Paulino-Afonso$^{1,6}$ \newauthor
Andra Stroe$^{8}$\thanks{ESO Fellow} \& Iv\'{a}n Oteo$^{5,8}$  \\ 
$^{1}$ Department of Physics, Lancaster University, Lancaster, LA1 4YB, UK \\
$^{2}$ Leiden Observatory, Leiden University, P.O.\ Box 9513, NL-2300 RA Leiden, The Netherlands \\
$^{3}$ Cahill Center for Astrophysics, California Institute of Technology, 1216 East California Boulevard, Pasadena, CA 91125, USA \\
$^{4}$ Centre for Extragalactic Astronomy, Department of Physics, Durham University, South Road, Durham DH1 3LE UK \\
$^{5}$ Institute for Astronomy, University of Edinburgh, Royal Observatory, Blackford Hill, Edinburgh EH9 3HJ, UK \\
$^{6}$ Departamento de F\'{i}sica, Faculdade de Ci\^{e}ncias, Universidade de Lisboa, Campo Grande, PT1749-016 Lisbon, Portugal \\
$^{7}$ Department of Physics and Astronomy, University of California, 900 University Ave., Riverside, CA 92521, USA \\
$^{8}$ European Southern Observatory, Karl-Schwarzschild-Str. 2, 85748 Garching, Germany }

\date{Accepted 2018 March 21. Received 2018 March 17; in original form 2018 February 27.}
\pubyear{2018}

\begin{document}
\label{firstpage}
\pagerange{\pageref{firstpage}--\pageref{lastpage}}
\maketitle

\begin{abstract}
Deep narrow-band surveys have revealed a large population of faint Ly$\alpha$ emitters (LAEs) in the distant Universe, but relatively little is known about the most luminous sources ($\rm L_{\rm Ly\alpha}\gsim 10^{42.7}$\,erg\,s$^{-1}$; $\rm L_{\rm Ly\alpha}\gsim L^*_{\rm Ly\alpha}$). Here we present the spectroscopic follow-up of 21 luminous LAEs at $z\sim2-3$ found with panoramic narrow-band surveys over five independent extragalactic fields ($\approx4\times10^6$\,Mpc$^{3}$ surveyed at $z\sim2.2$ and $z\sim3.1$). We use WHT/ISIS, Keck/DEIMOS and VLT/X-SHOOTER to study these sources using high ionisation UV lines. Luminous LAEs at $z\sim$\,2--3 have blue UV slopes ($\beta=-2.0^{+0.3}_{-0.1}$), high Ly$\alpha$ escape fractions ($50^{+20}_{-15}$\%) and span five orders of magnitude in UV luminosity ($\rm M_{UV}\approx-19$ to $-24$). Many (70\%) show at least one high ionisation rest-frame UV line such as C{\sc iv}, N{\sc v}, C{\sc iii}], He{\sc ii} or O{\sc iii}], typically blue-shifted by $\approx100-200$\,km\,s$^{-1}$ relative to Ly$\alpha$. Their Ly$\alpha$ profiles reveal a wide variety of shapes, including significant blue-shifted components and widths from 200 to 4000\,km\,s$^{-1}$. Overall, $60\pm11$\,\% appear to be AGN dominated, and at $\rm L_{\rm Ly\alpha}>10^{43.3}$\,erg\,s$^{-1}$ and/or $\rm M_{UV}<-21.5$ virtually all LAEs are AGN with high ionisation parameters ($\log U=0.6\pm0.5$) and with metallicities of $\approx0.5-1$\,$Z_{\odot}$. Those lacking signatures of AGN ($40\pm11$\,\%) have lower ionisation parameters ($\log U=-3.0^{+1.6}_{-0.9}$ and $\log\xi_{\rm ion}=25.4\pm0.2$) and are apparently metal-poor sources likely powered by young, dust-poor ``maximal" starbursts. Our results show that luminous LAEs at $z\sim$\,2--3 are a diverse population and that $2\times$\,L$^*_{\rm Ly\alpha}$ and $2\times$\,M$_{\rm UV}^*$ mark a sharp transition in the nature of LAEs, from star formation dominated to AGN dominated.
 \end{abstract}


\section{Introduction}

There has been a steady increase in the number of sources identified at moderately high redshifts ($z\gsim$\,2--3) over the past twenty years (e.g. \citealt{Ellis2013,Bouwens2014,Bouwens2015,Bowler2014,Finkelstein2015,Atek2015,Santos2016}). Most have been found using the Lyman-break technique \citep[e.g.][]{Koo1980,Guhathakurta1990,Steidel1993,Giavalisco1996}, combined with deep, multi-band imaging with the {\it Hubble Space Telescope} ({\it HST}). However, spectroscopy of faint continuum-selected candidates has progressed at a much slower pace as spectroscopic continuum detections are difficult for such faint sources. A different approach is to find and study sources with bright Ly$\alpha$ in emission \citep[e.g.][]{Ouchi2008,Zitrin2015,Oesch2015,Sobral2015,Hu2016,Matthee2017c,Jiang2017} across a range of redshifts \citep[e.g.][]{Moller1993,Harikane2017,Sobral2017_SC4K}. While these can now be found with wide-field ground-based surveys \citep[e.g.][]{Matthee2015,Hu2016,Santos2016,Shibuya2017,Zheng2017}, much is unknown about their nature, metallicity and stellar populations \citep[see e.g.][]{Wold2014,Nilsson2007,Matthee_CR7_17,Sobral2017_SC4K}.

Significant spectroscopic progress has also been made by targeting lensed galaxies \citep[e.g.][]{Swinbank2010,Vieira2013,Stark2014,Stark2015}, allowing the detection and study of intrinsically very faint lines. Combining results from intrinsically bright and faint sources reveals a picture in which galaxies appear to have ubiquitous high equivalent width nebular emission lines at high redshift \citep[e.g.][]{Smit2014,Sobral2014,Smit2016,MArmolQueralto2016,Khostovan2016}. This may be due to high redshift galaxies producing more ionising photons per UV luminosity \citep[e.g.][]{Topping2015,Bouwens2015X,Matthee2017,Stanway2017}. Recent results also highlight that high redshift galaxies can have hard ionising spectra, as indicated by UV emission lines such as C{\sc iii}], C{\sc iv}, O{\sc iii}] or He{\sc ii} \citep[e.g.][]{StarkCIV,Sobral2015,Jaskot2016,Schmidt2017,Laporte2017}. This is in agreement with more detailed, larger statistics results, which show an increase in the ionisation parameter with redshift, traced by e.g. [O{\sc iii}]/[O{\sc ii}] \citep[e.g.][]{Erb2010,NakajimaOuchi2013,Khostovan2016,Nakajima2016,Vanz2016,Vanz2017} or evidence for hard spectra from iron-poor stellar populations \citep[][]{Steidel2016}. Understanding the origin of the strong evolution relative to local galaxies is still a major open question. Many mechanisms/physical processes have been proposed, including binary stars, bursty star formation histories, active galactic nuclei and several effects at low metallicity \citep[e.g.][]{Eldridge2008,Eldridge2009,Eldridge2012,Ma2016,Grafener2015,Stanway2016}.

Furthermore, the common approach of selecting Lyman-break galaxies has resulted in statistically large samples, but failed to reveal a convincing population of Lyman continuum (LyC) leakers \citep[e.g.][]{Siana2007,Guaita2016,Marchi2017,Marchi2017b}, required for re-ionisation \citep[e.g.][]{Faisst2016}. This may be a consequence of selection, as Lyman-break galaxies (LBGs) are biased against sources leaking significant amounts of LyC photons \citep[e.g.][]{Cooke2014}. LBGs tend to be more evolved and show little to no Ly$\alpha$ in emission \citep[e.g.][]{Cassata2015}. In addition, recent theoretical and observational work suggest an important link between LyC and Ly$\alpha$ escape fractions \citep[e.g.][]{Verhamme2015,Verhamme2017,Dijkstra2016,Izotov2017_Lya,Vanzella2018}, implying that Ly$\alpha$ emitters (LAEs) contribute significantly to the LyC luminosity density. The faint end slope of the Ly$\alpha$ luminosity function (LF) is steep ($\alpha\sim-2$), implying a large number of faint LAEs \citep[e.g.][]{Rauch2008,Gronke2015,Dressler2015,Santos2016,Drake2017_AeA} and LAEs produce typically many more ionising photons per UV luminosity than LBGs \citep[e.g.][]{Nakajima2016,Matthee2017,Harikane2017}. The combination of recent results on LAEs is particularly promising and further enhances the motivation to study those: the high equivalent width (EW) LAEs may be exactly what is needed to re-ionise the Universe, having ``high'' escape fractions \citep[e.g.][]{Verhamme2017}, high production of ionising photons \citep[e.g.][]{NakajimaOuchi2013,Nakajima2016,Nakajima2018}, and large number densities \citep[e.g.][]{Drake2017_AeA,Sobral2017_SC4K}.

Despite recent progress, little is known about the nature and redshift evolution of the most luminous LAEs ($\rm L_{\rm Ly\alpha}\sim10^{43-44.5}$\,erg\,s$^{-1}$). At $z\sim$\,2--3 spectroscopic studies have mostly focused on the follow-up of fainter, more numerous LAEs typically below the characteristic Ly$\alpha$ ($\rm L_{\rm Ly\alpha}\lesssim10^{43}$\,erg\,s$^{-1}$) and UV luminosities at that redshift \citep[e.g.][]{Fynbo2001,Fynbo2003,Ouchi2008,Grove2009,Trainor2015,Hathi2016_VUDS}, or by blindly finding and studying faint LAEs with deep but small volume searches with IFU instruments such as MUSE \citep[e.g.][]{Drake2017,Hashimoto2017,Caruana2018}. Important progress has also been made by studying extremely rare quasars found with SDSS \citep[e.g.][]{Richards2006}, $\approx7-10$ orders of magnitude brighter in the UV and revealing large Ly$\alpha$ haloes with $\rm L_{\rm Ly\alpha}\sim10^{43-44}$\,erg\,s$^{-1}$ \citep[e.g.][]{Borisova2016}. Still, little is known about the general population of bright LAEs, although there is some evidence for these sources to have a significant AGN contribution \citep[e.g.][]{Ouchi2008,Nilsson2009,Sobral2017,Matthee2017b}.

In this paper we study bright LAEs at $z\sim$\,2--3 through relatively deep spectroscopic follow-up, in order to unveil their nature and physical properties. These bright sources span the parameter space between faint LAEs and rare quasars. The paper is organised as follows. In \S\ref{obs_data} we present the observations of our sample of bright Ly$\alpha$ candidates with WHT/ISIS, Keck/DEIMOS and VLT/X-SHOOTER, followed by the respective data reduction. In \S\ref{measurements_analysis} we present the measurements and analysis. Results are presented in \S\ref{results}. The nature of bright LAEs is studied in \S\ref{UV_lines_nat}, based on high ionisation UV lines. We discuss our results in \S\ref{discussion}. Conclusions are presented in \S\ref{conclusions}. Throughout this paper, we use AB magnitudes \citep[][]{Oke1983}, a Salpeter \citep[][]{Salpeter1955} IMF and a $\Lambda$CDM cosmology with $H_0=70$\,km\,s$^{-1}$\,Mpc$^{-1}$, $\Omega_{\rm M}=0.3$ and $\Omega_{\Lambda}=0.7$.

\section{Observations and Data Reduction} \label{obs_data}

\subsection{Sample selection}

\begin{figure}
\includegraphics[width=8.5cm]{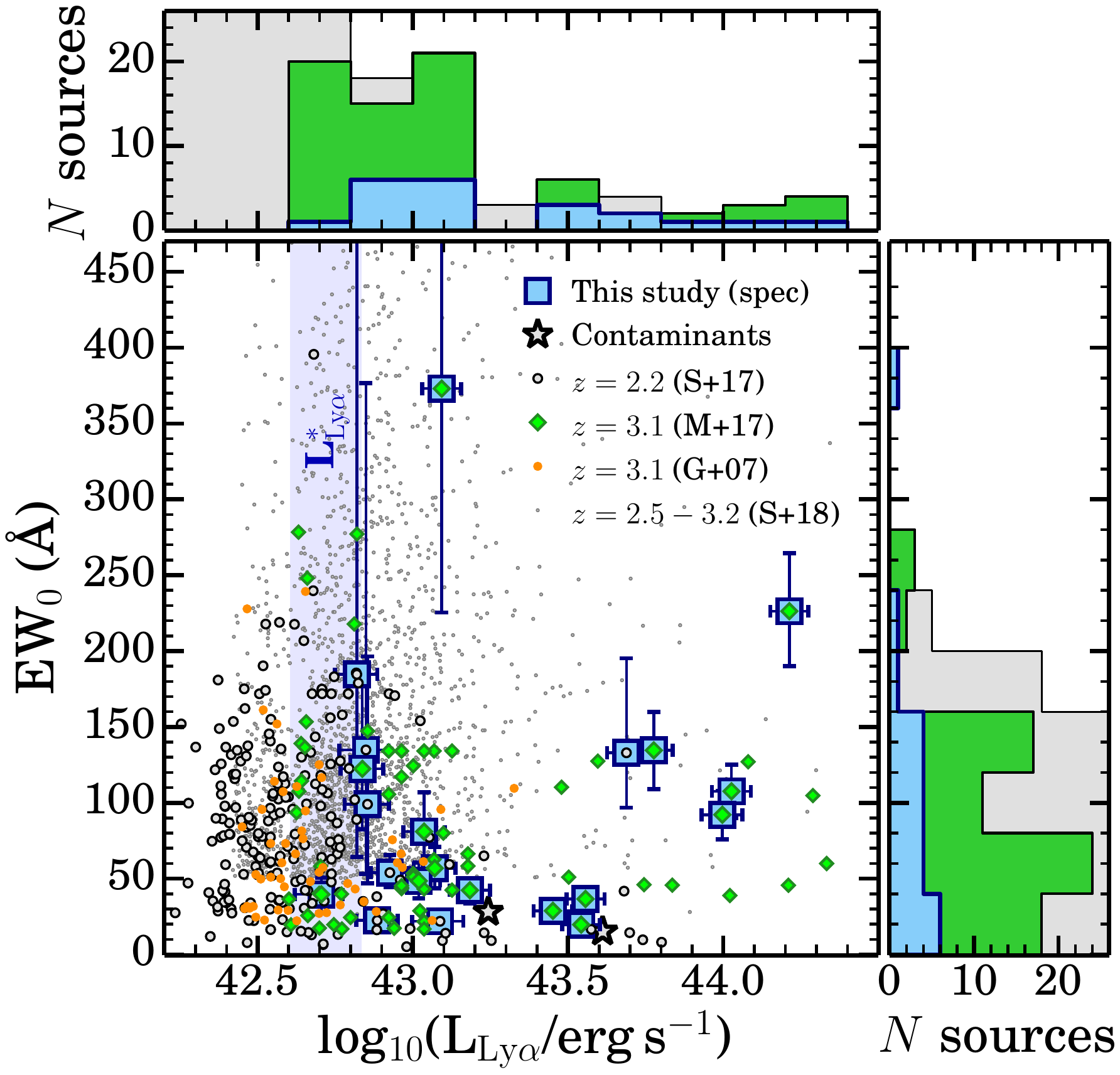}
\caption{Comparison between the distribution of Ly$\alpha$ luminosities and rest-frame EWs (both quantities derived from narrow-band imaging) of the $z=2.2$ \citep[from the CALYMHA survey, S+17;][]{Sobral2017,Matthee2017b} and $z=3.1$ parent samples \citep[M+17;][]{Matthee2017b}. We also show a sample of fainter LAEs at $z=3.1$ probing a smaller volume \citep[G+07;][]{Gronwall2007} and the large sample of Ly$\alpha$ emitters at $z\sim2.5-3.2$ \citep[SC4K;][]{Sobral2017_SC4K}. L$^*_{\rm Ly\alpha}$ (the knee/typical luminosity of the Ly$\alpha$ luminosity function; \citealt{Sobral2017_SC4K}) is indicated at $z\sim$\,2--3 as a shaded region, representing the errors and positive evolution from $z\sim2$ to $z\sim3$. Our spectroscopic sample is representative of L$_{\rm Ly\alpha}\gsim {\rm L^*_{\rm Ly\alpha}}$ LAEs, probing a large range in luminosities and EWs.}
\label{Lum_EW_parent}
\end{figure}

Our luminous LAEs are selected from wide-field narrow-band Ly$\alpha$ surveys. Candidate bright LAEs at $z\sim2.2$ are selected from the 1.43\,deg$^2$ CALYMHA survey \citep[EW$_0>5$\,{\AA}, covering the UDS and COSMOS fields, see][]{Matthee2016,Sobral2017,Sobral2017_SC4K}. We select those with Ly$\alpha$ luminosities above 10$^{42.8}$\,erg\,s$^{-1}$ (up to 10$^{43.6}$\,erg\,s$^{-1}$); see Figure \ref{Lum_EW_parent}. Luminous $z\sim3.1$ Ly$\alpha$ candidates are selected from a total of $\sim3.1$\,deg$^2$ over the GOODS-N, SA22 and Bo\"{o}tes fields \citep[EW$_0>25$\,{\AA}; see][]{Matthee2017b} to have luminosities in excess of 10$^{42.7}$\,erg\,s$^{-1}$ (up to 10$^{44.2}$\,erg\,s$^{-1}$). Our targets have number densities of 10$^{-4}$--10$^{-6}$\,Mpc$^{-3}$ \citep[e.g.][]{Matthee2017b,Sobral2017_SC4K}, and have been found in a total volume of $\approx4\times10^6$\,Mpc$^{3}$. We note that due to the selection applied to obtain candidate LAEs from the full sample of emission line candidates \citep[see][]{Sobral2017_SC4K} our samples will be biased against potential (galaxy-galaxy) lensed sources, as the continuum of the foreground lens will lead to classifying lensed LAEs as low redshift interlopers. 

In total, we have targeted 23 sources as candidate luminous LAEs. We present their Ly$\alpha$ luminosities and rest-frame Equivalent Widths (EW$_0$) in Figure \ref{Lum_EW_parent}, compared to the parent sample of more typical, lower luminosity LAEs at both $z\sim2.2$ \citep[][]{Sobral2017} and $z\sim3.1$ \citep[][]{Matthee2017b}. As Figure \ref{Lum_EW_parent} shows, our sample is representative of the most luminous LAEs, covering 1.5\,dex in Ly$\alpha$ luminosities (10$^{42.7-44.2}$\,erg\,s$^{-1}$) and 1.2\,dex in rest-frame Ly$\alpha$ EW$_0$ ($20-400$\,\AA).

%
%

\subsection{Spectroscopic observations}

\subsubsection{WHT/ISIS}

We used the Intermediate dispersion Spectrograph and Imaging System (ISIS) on the William Herschel Telescope (WHT) to observe $z\sim$\,2--3 Ly$\alpha$ candidates. ISIS is a double-armed, medium-resolution spectrograph, which allows simultaneous observing with the blue and red arms.

A total of six $z\sim$\,2--3 Ly$\alpha$ candidates were selected from GOODS-N and Bo\"{o}tes. We observed each source for roughly one hour during May 4-6 and Jun 30-Jul 01 2016; see Table \ref{Obs_table_ISIS}. All observations were done under clear conditions and the seeing varied between $\sim0.4''$ and $\sim1.5''$ (typical seeing $\approx0.9''$). We used a 1$''$ slit for all observations.

The blue and red arms were used with the R600 gratings, with central wavelengths 4800\,{\AA} ($R\sim2400$) and 7000\,{\AA} ($R\sim3900$) on the first three nights and with the R300B/R316R gratings with central wavelengths 5000\,{\AA} ($R\sim1200$) and 8000\,{\AA} ($R\sim2200$) for the final two nights. We obtained biases, arcs and flats for each grating, arm and central wavelength, at the start and end of each night.

Our targets are all too faint (typically $I\sim23-24$\,mag) in the continuum to be directly acquired using WHT/ISIS, and thus we identified a relatively nearby star ($I\sim13-16$\,mag, $\sim20-60''$ away from our target) which we used for acquisition. We conduct our observations by first acquiring the offset star, taking a 30-60\,s (depending on its magnitude) exposure on the star, then blindly offsetting the telescope to the target and taking one 900-1000\,s exposure of the target. After each science exposure, we go back to the star, offset along the slit by $\pm10''$, acquire the star again, take another star spectrum, and then offset to the science target to take another exposure. This procedure allows us to use the star as a flux calibrator and slit loss estimator. Our procedure also allows us to extract the trace and its curvature in an optimal way, which we use to combine and extract the spectra (see \S\ref{WHIS_red}).

\subsubsection{Keck/DEIMOS}

DEIMOS was used to spectroscopically target eight luminous Ly$\alpha$ candidates at $z\sim3$ in five different masks for the Bo\"{o}tes field, and in one mask for GOODS-N over three different nights in June 2, July 6 and July 29 in 2016. We have also observed other line emitters; see Table \ref{Obs_table_DEIMOS} and \citealt{Matthee2017b}. Observations were conducted under clear conditions. The seeing was in the range $\approx0.6''-0.9''$. We used a central wavelength of 7200\,{\AA} and the 600L grating ($R\sim2400$), with a pixel scale of 0.65\,\AA\,pix$^{-1}$, which allowed us to probe from 4550\,\AA \ to 9850\,\AA. We used the 0.75$''$ slit, in the same mode as used by \cite{Darvish2015}.

We obtained biases, arcs and flats at the start of each observing night. Typical science exposures of 900\,s were obtained with a $\pm2.5''$ dithering for two of the nights and without dithering for the final observing night on 30 July 2016. Observations are listed in Table \ref{Obs_table_DEIMOS}.

\subsubsection{VLT/X-SHOOTER}

We used X-SHOOTER \citep{Vernet2011} to observe eleven candidate luminous $z=2.2-3.1$ LAEs in October 2016, January, June-July, December 2017 and January 2018 (see Table \ref{Obs_table_XSHOOT}). X-SHOOTER allows to simultaneously obtain a relatively high resolution spectrum with the UV/blue (UVB), visible (VIS) and near-infrared (NIR) arms, providing a coverage from 3000 to 24,800\,\AA. The seeing varied between 0.8$''$ and 1.5$''$ (median seeing $0.9''$) and observations were done under clear conditions. We used the low read-out speed without binning. Typical exposure times of 200\,s, 300\,s and $4\times80$\,s in the UV, optical and NIR arms were used, respectively; see Table \ref{Obs_table_XSHOOT}. We first acquired a star (with $I$-band magnitudes 16-17 AB) and applied a blind offset to the target. We nodded along the slit from an A to a B position (typically 4-6$''$ apart), including a small jitter box in order to always expose on different pixels. We used 1\,$''$ slits for both the optical and near-infrared arms (resolution of $R\sim2500$ and $R\sim4400$, for the optical and near-infrared arms, respectively) and the 1.2\,$''$ slit in the UV ($R\sim4400$). The total exposure times for each source are given in Table \ref{Obs_table_XSHOOT}.

\subsection{Data Reduction}

\subsubsection{WHT/ISIS: the William Herschel ISIS pipeline}\label{WHIS_red}

We developed a {\sc python} pipeline to reduce ISIS spectra\footnote{The William Herschel ISIS pipeline (see Appendix \ref{WHIS_pipeline}).} and exploit our observing setup/methodology. We reduce the data on a night by night basis, and only combine data from different nights for extracted 1D spectra. Our pipeline follows standard steps: we bias-subtract, flat-field and wavelength-calibrate the spectra using appropriate calibration frames. Additional wavelength calibration is performed by exploiting the sky lines in deep spectroscopic exposures. We describe the processes in more detail in Appendix \ref{WHIS_pipeline}. In Figure \ref{2D_spectra_WHTIS} we show our reduced 2D spectra, centred on the Ly$\alpha$ line within a window of $\pm4000$\,km\,s$^{-1}$.

We flux calibrate our spectra by using the blind offset stars and their SDSS magnitudes. The resolution (FWHM) of each arm/grating is estimated with un-blended sky lines and found to be 2 and 1.8\,{\AA} for the R600B/R600R gratings spectra and 4.0 and 3.6\,{\AA} FWHM for the R300B/R316R gratings. We bin our 1D spectra by roughly one third of the corresponding FWHM, which, in practice, results in binning by 2 spectral pixels. We show the typical rms per binned resolution element in Table \ref{noise_resolution_final}.

%
%
\begin{figure}
\includegraphics[width=8.29cm]{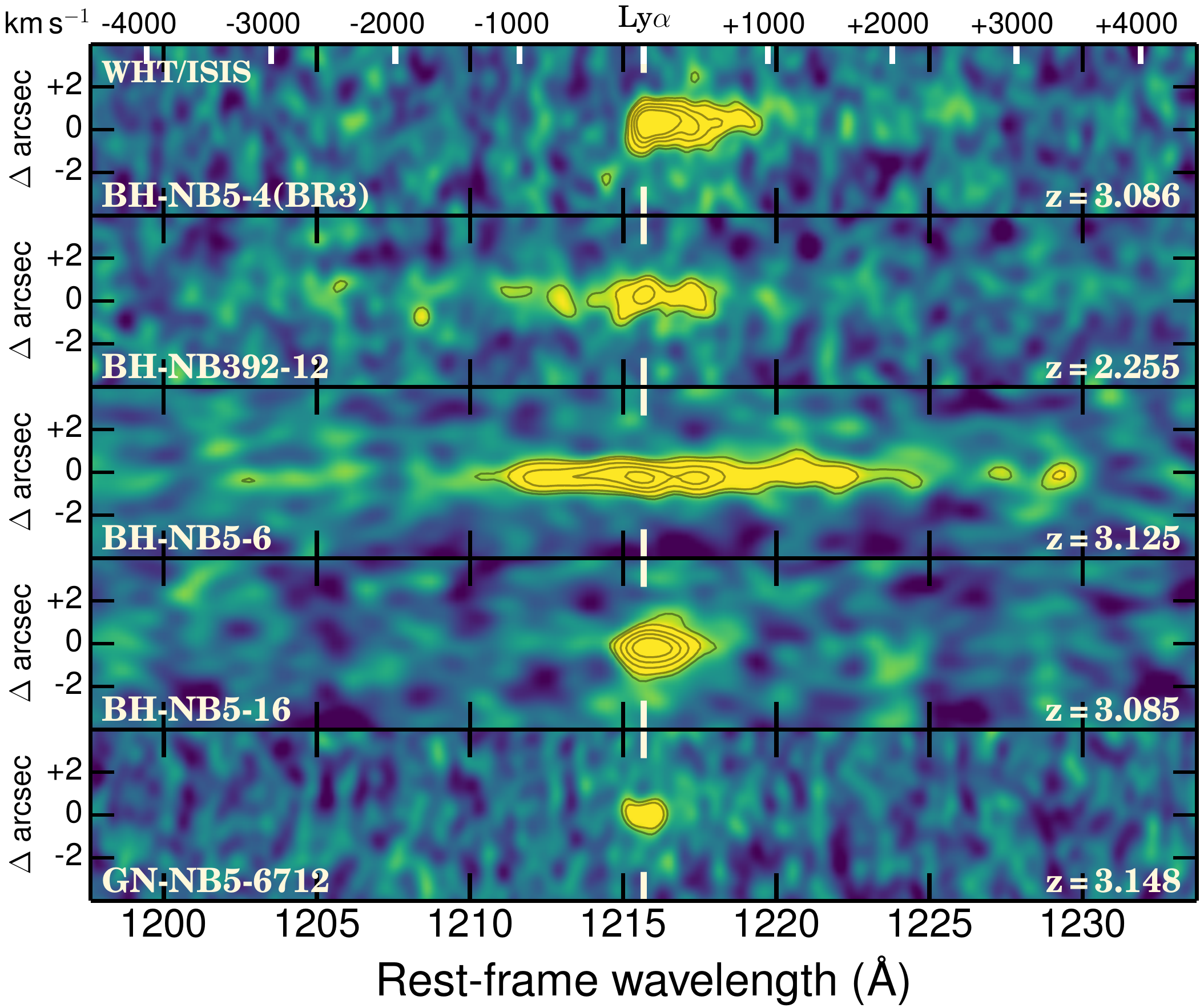}
\caption{The reduced 2D Ly$\alpha$ spectra, displayed in the S/N space, smoothed with a Gaussian kernel of 2 pixels, from our WHT/ISIS observations, labelled with ID and redshift, and ordered by Ly$\alpha$ flux top to bottom (brightest on top). We show $1.5,2,3,4,5$\,$\sigma$ contours and use contrast cutoffs at $-1$ and $+2$\,$\sigma$ to display the 2D spectra. The top axis shows the velocity shifts from the Ly$\alpha$ redshift. We find one broad-line emitter (BH-NB5-6), and one with a blue wing/complicated dynamics (BH-NB392-12). The brightest source observed with WHT/ISIS is BR3, which shows the most asymmetric Ly$\alpha$ line profile. The faintest source targeted with Ly$\alpha$ coverage is GN-NB5-6712, which is also one of the sources showing the narrowest Ly$\alpha$ profile.}
\label{2D_spectra_WHTIS}
\end{figure} 

\subsubsection{Keck/DEIMOS: spec2d pipeline}

All Keck/DEIMOS data were reduced using the DEIMOS {\sc spec2d} pipeline \citep{Cooper12}, which follows very similar steps to our WHT/ISIS data reduction. Briefly, each observed spectra were flat-fielded, cosmic-ray-removed, sky-subtracted and wavelength-calibrated \citep[see][]{Darvish2015}. We used standard Kr, Xe, Ar and Ne arc lamps for the wavelength calibration. We also checked the wavelength solution at the end of the pipeline reduction, taking advantage of the numerous OH sky lines. For the observations obtained with a dithering pattern, we find significant offsets between the observed sky line positions and the correct OH line wavelengths of over 10-20\,\AA, varying in a non-linear way in the blue and in the red. Therefore, for all observations for which we dithered we identify $\sim100$ unblended OH lines and obtain a final wavelength calibration, which produces an rms of $\approx0.5$\,{\AA}, about $\sim6-7$ times better than the resolution.

\begin{figure*}
\begin{tabular}{cc}
\includegraphics[width=8.1cm]{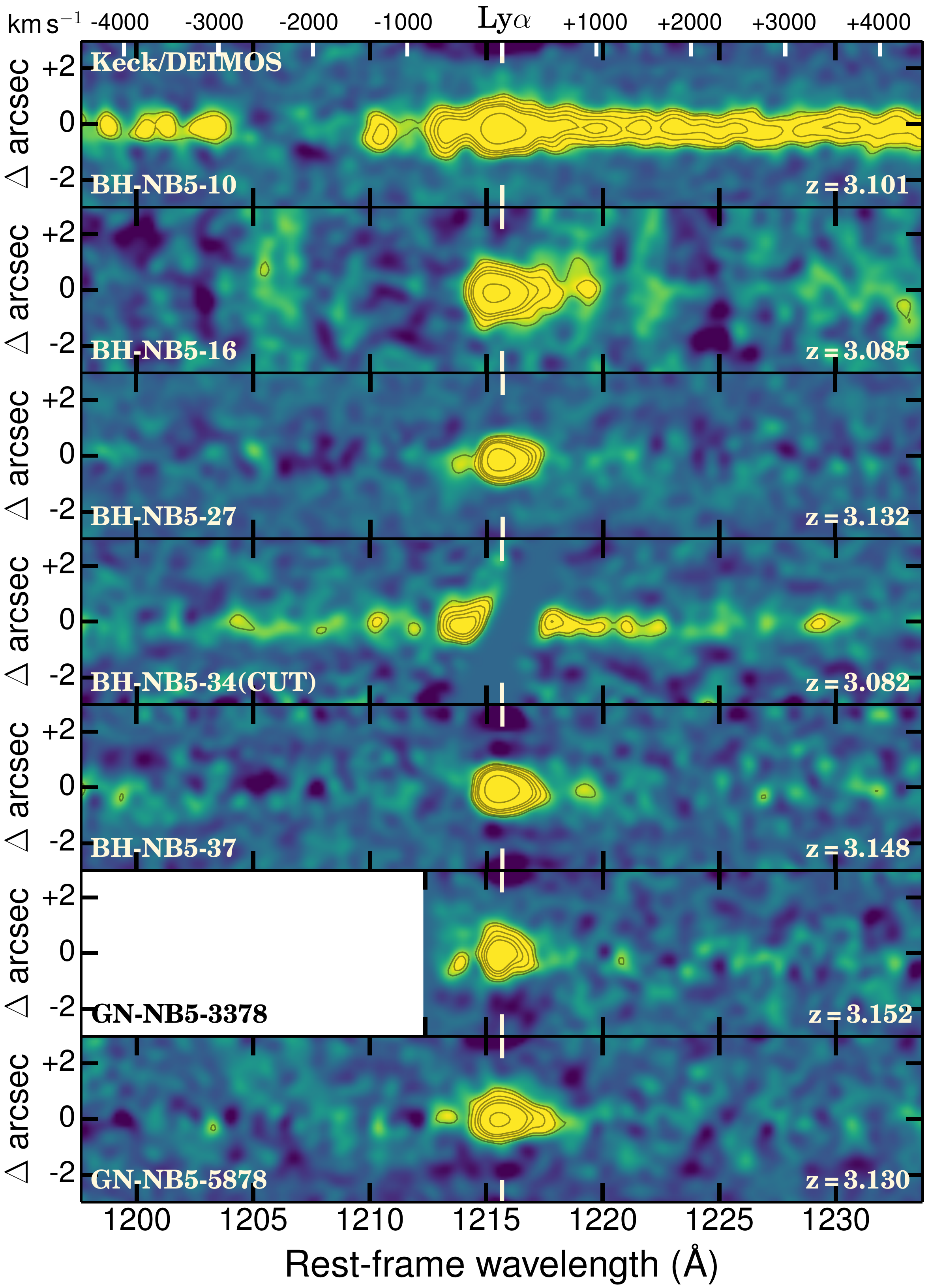}&
\includegraphics[width=7.95cm]{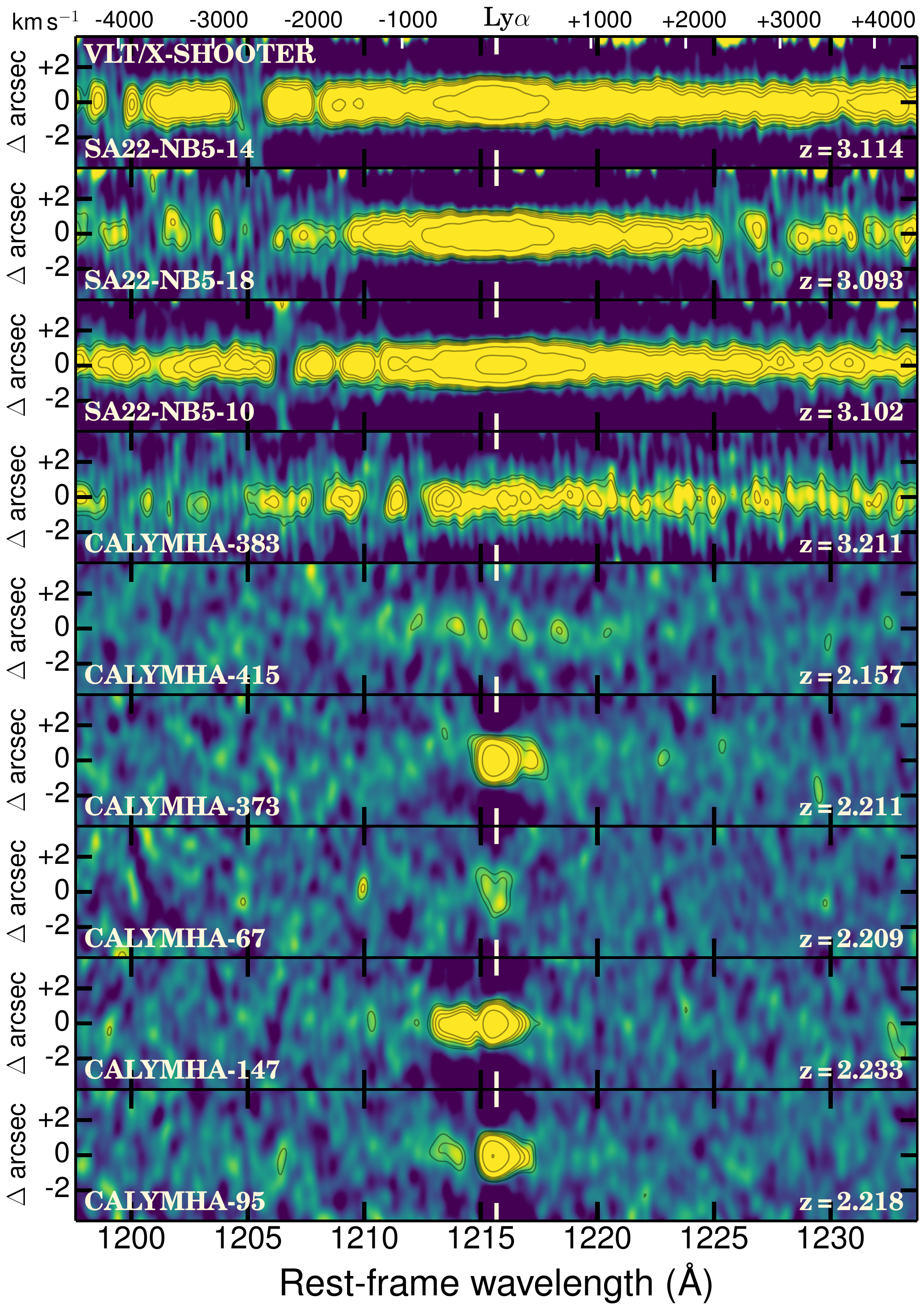}\\
\end{tabular}
\caption{The reduced 2D spectra of luminous LAEs from our DEIMOS ({\it left}) and X-SHOOTER ({\it right}) observations, smoothed with a Gaussian kernel of 2 pixels, labelled with ID and redshift, listed from the brightest (top) to the faintest (bottom) per instrument. We show $1.5,2,3,4,5,10,20$\,$\sigma$ contours. Our LAEs present a wide variety of Ly$\alpha$ line profiles with some broad lines (FWHMs in excess of 1000\,km\,s$^{-1}$) at the highest Ly$\alpha$ luminosities, and narrower Ly$\alpha$ lines at lower Ly$\alpha$ luminosities. Five sources show a blue-shifted Ly$\alpha$ component, with CALYMHA-147 being the most noticeable example and showing the highest blue-to-red ratio ($\approx0.6\pm0.1$). LAEs with significant UV continuum also reveal the Ly$\alpha$ forest. Note that the ``negative flux'' regions are a consequence of jittering along the slit.}
\label{2D_spectra_DEI_XSHOOT}
\end{figure*}

For the final night of observations with Keck/DEIMOS, no dithering pattern was used for sky subtraction. Without dithering, the wavelength calibration provided by the pipeline was found to be accurate within 0.5\,{\AA}, and no further correction was necessary. We show the reduced 2D spectra in Figure \ref{2D_spectra_DEI_XSHOOT}.

The pipeline also generates the 1D spectrum extraction from the reduced 2D spectrum, following the optimal extraction algorithm of \cite{Horne1986}. This extraction creates a one-dimensional spectrum of the target, containing the summed flux at each wavelength in an optimised window. We flux calibrate the data with bright enough sources within the masks, and also observations of the Feige 66 standard star. We bin our spectra to one third of the observed resolution (FWHM) of 3\,{\AA} (which results in binning 2 spectral pixels, $\approx1$\,{\AA}). Our final resolution binned spectra have a typical rms per binned resolution element (1\,$\sigma$) of $3-5\times10^{-19}$\,erg\,s$^{-1}$\,cm$^{-2}$\,{\AA}$^{-1}$ (see Table \ref{noise_resolution_final}) and thus typically a factor $\sim10$ deeper than WHT/ISIS.

%
%

\subsubsection{VLT/X-SHOOTER: ESO pipeline}

We use the ESO X-SHOOTER pipeline \citep{Modigliani2010} to reduce the UVB, VIS and NIR spectra separately \citep[see also][]{Matthee2017c}. The data reduction steps follow closely those implemented for both ISIS and DEIMOS data, with the necessary differences, particularly for the X-SHOOTER NIR arm which allow to obtain [O{\sc iii}] and H$\alpha$ for the $z\sim2.2$ LAEs (NIR spectral properties will be fully discussed in Matthee et al. in prep.).

In short, we start by grouping frames. We then identify and process the bias frames (to create master biases in the UVB/VIS arms) and dark current (NIR arm). We produce master flats per arm and flat-field the data. Finally, sky subtraction and wavelength calibration are done. In order to be able to flux calibrate the spectra, several standard stars have been observed (see Table \ref{Obs_table_XSHOOT}). We use the X-SHOOTER pipeline to reduce the standard stars in the same way as the science targets and combine the exposures from single observing blocks.

In the case that a source has been observed by multiple observing blocks, we co-add the frames by weighting each frame with the inverse of the variance (noise) in 2D. Corrections for slight positional variations based on the position of the peak of observed Ly$\alpha$ lines were applied. We show the reduced 2D spectra in Figure \ref{2D_spectra_DEI_XSHOOT}, ordered by their narrow-band-derived Ly$\alpha$ luminosity.

We extract 1D spectra by summing the counts in an optimised spatial window, which typically corresponds to a total of $\sim10$ spatial pixels (1.8$''$). We use isolated sky lines together with arc lines to find that the resolution of our spectra (FWHM) is 1.6\,{\AA} in the UVB and VIS arms and 3.6\,{\AA} in the NIR arm. We finally bin our spectra to one third of the FWHM which, in practice, corresponds to binning 3 and 2 spectral pixels for the UVB/VIS and NIR arms, respectively. Table \ref{noise_resolution_final} provides the typical rms per binned resolution element for our binned spectra.

\section{Measurements and Analysis} \label{measurements_analysis}

\subsection{Spectroscopic redshifts and line identification}


All our spectra are calibrated to air wavelengths. Before we determine redshifts, we first convert from air wavelengths to vacuum \citep[see][]{Morton1991}, and then use vacuum rest-frame wavelengths for the rest of the paper (see Table \ref{table_lines}).

We start by obtaining a redshift using the Ly$\alpha$ emission only ($z_{\rm Ly\alpha}$), and identifying its peak (see Figures \ref{2D_spectra_WHTIS}, \ref{2D_spectra_DEI_XSHOOT} and \ref{1D_examples_fitting}). We store these as the redshift of the Ly$\alpha$ emission line, and use those as a first approximation of the redshift of the source. We then look for other lines (see Table \ref{table_lines}) by searching and fitting Gaussians (or double Gaussians) and deriving a redshift for them, allowing for velocity offsets up to $\pm1000$\,km\,s$^{-1}$.

%
%
\begin{table}
 \centering
\caption{A list of high ionisation rest-frame UV and optical lines used in this paper; see also \citealt{Veilleux2002}. We list them in vacuum wavelengths \citep[see][]{Morton1991}. For spectra with high enough resolution to resolve doublet lines we also attempt to fit a double Gaussian. For simplicity, and unless the S/N is high enough, we fix the doublet separation and impose that both lines should have the same FWHM.}
 \label{table_lines}
\begin{tabular}{cccc}
\hline
Emission   & $\lambda_{\rm vacuum}$  & Ionisation & \# Detections \\ 
line  &   (\AA) &  Energy (eV) & (99.4\% conf.)\\ 
\hline
Ly$\alpha$ & 1215.67 & 13.6 & 20 (100\%) \\
N{\sc v} &  1238.8,1242.8 & 77.4 & 6 (33\%) \\
O{\sc iv]} & 1401,1407 & 54.9 & 1 (11\%) \\ 
N{\sc iv]} & 1483.4,1486.6 & 47.4 & 2 (11\%) \\
C{\sc iv} &  1548.2,1550.8 & 47.9 & 8 (40\%) \\
He{\sc ii} &  1640.47 & 54.4 & 5 (25\%) \\
O{\sc iii]} &  1661,1666 & 35.1 & 1 (5\%)  \\ 
N{\sc iii]} &  1749.7,1752.2 & 29.6 & 2 (11\%) \\
C{\sc iii]} &  1907,1910 & 24.4 & 4 (25\%) \\
\hline
\end{tabular}
\end{table}

\subsection{Contaminants}\label{contaminants}

We confirm that the vast majority (91\%) of our 23 candidates are luminous LAEs, but we also find two contaminants. These include one low redshift [O{\sc ii}] emitter at $z=0.056$ (CALYMHA-438) and a star with a large number of absorption bands (CALYMHA-85); both with very low EWs (see Figure \ref{Lum_EW_parent}). These imply an overall contamination of $\approx10$\,\%, likely lower at $z=3.1$ than at $z=2.23$ (a consequence of the very low EW cut used for $z=2.23$; see discussion in \citealt{Sobral2017}). We further study the potential contamination and completeness of the larger sample of bright narrow-band selected sources from the parent samples \citep[][]{Sobral2017,Matthee2017b} by using the follow-up of other sources with Keck/DEIMOS \citep[see][]{Matthee2017b}. Based on one missed real LAE and 2 extra [O{\sc ii}] contaminants \citep[see][for further details]{Matthee2017b}, we estimate a conservative contamination of $\approx16$\,\% (four contaminants in a sample of 25 sources) at $z\sim$\,2--3 for our parent sample of bright LAEs and a completeness of $>90$\% (only one real LAE missed by our selection).

%
%
%
\begin{table*}
 \centering
  \caption{Our spectroscopic sample of luminous LAEs, ranked by redshift and then by Ly$\alpha$ luminosity of the parent sample (high to low luminosity top to bottom). We present the main measured photometric and Ly$\alpha$ properties and also provide the most likely nature classification of each source, based on the FWHM of the Ly$\alpha$ line (BL-AGN vs NL emitters), but also on the ratios between the different rest-frame UV emission lines for the NL emitters (AGN vs SF; see \S\ref{NL_AGN_iden}). All errors are the 16th and 84th percentiles. Some uncertainties in the Ly$\alpha$ redshift are lower than 0.001, but here we set them 0.001 if they are below that value. We add 15\% uncertainties in the flux calibration or zero-point calibration in quadrature for the final error on the Ly$\alpha$ luminosity. (1) CALYMHA-147 shows a bright blue-component and the FWHM given is a fit to both components; the FWHM of the red component is $\approx180$\,km\,s$^{-1}$, while the blue component has a FWHM of $\approx400$\,km\,s$^{-1}$. (2) The FWHM of this source should be interpreted with care due to the lack of coverage towards the redder part of Ly$\alpha$ (see e.g. Figure \ref{2D_spectra_DEI_XSHOOT}). This table is available on-line as a {\sc fits} catalogue without digit truncation.}
  \begin{tabular}{@{}ccccccccccc@{}}
  \hline
   ID & R.A.  & Dec.  &  $z_{\rm spec}$ & M$_{\rm UV}$  & $\beta_{UV}$   & $\log$\,L$_{\rm Ly\alpha}$  & EW$_0$ & FWHM & Class.    \\
(This paper)  & (J2000)  & (J2000)   & (Ly$\alpha$)  & (AB) & & (erg\,s$^{-1}$) & (\AA) &   (km\,s$^{-1}$) &  \\
 \hline
   \noalign{\smallskip}
 BH-NB392-12  & 14\,30\,28.55  & $+33\,33\,29.0$  & $2.255^{+0.001}_{-0.001}$ &  $-21.9^{+0.2}_{-0.2}$   &  $-1.4^{+0.2}_{-0.2}$   &  $43.69^{+0.06}_{-0.06}$   &  $130^{+50}_{-40}$  &  $1320^{+240}_{-200}$   & BL-AGN  \\
 CALYMHA-415  & 02\,16\,33.29  & $-05\,17\,57.8$  & $2.153^{+0.004}_{-0.003}$ &  $-21.1^{+0.2}_{-0.2}$   &  $-2.0^{+0.1}_{-0.1}$   &  $43.18^{+0.07}_{-0.07}$   &  $22^{+5}_{-4}$  &  $2010^{+450}_{-570}$   & BL-AGN  \\
 BH-NB392-55  & 14\,30\,40.31  & $+34\,03\,20.6$  & $2.197^{+0.568}_{-0.003}$ &  $-22.4^{+0.2}_{-0.2}$   &  $-2.0^{+0.1}_{-0.1}$   &  $43.09^{+0.08}_{-0.08}$   &   $22^{+7}_{-6}$  & ---   & Unclass.  \\
 CALYMHA-373  & 02\,17\,46.13  & $-05\,02\,55.5$  & $2.211^{+0.001}_{-0.001}$ &  $-20.9^{+0.2}_{-0.2}$   &  $-2.1^{+0.2}_{-0.2}$   &  $42.93^{+0.06}_{-0.06}$   &  $54^{+11}_{-10}$  &  $260^{+20}_{-20}$   & NL-SF  \\
 CALYMHA-147  & 10\,00\,13.91  & $+01\,39\,24.3$  & $2.232^{+0.001}_{-0.001}$ &  $-19.9^{+0.2}_{-0.2}$   &  $-1.9^{+0.4}_{-0.4}$   &  $42.85^{+0.07}_{-0.08}$   &  $100^{+80}_{-40}$  &  $620^{+60}_{-90}$(1)   & NL-SF  \\
 CALYMHA-67  & 10\,01\,36.21  & $+02\,15\,16.8$  & $2.209^{+0.001}_{-0.001}$ &  $-19.7^{+0.2}_{-0.2}$   &  $-2.1^{+0.3}_{-0.3}$   &  $42.85^{+0.07}_{-0.08}$   &  $130^{+200}_{-70}$  &  $280^{+70}_{-60}$   & NL-SF  \\
 CALYMHA-95  & 10\,01\,06.55  & $+01\,45\,45.5$  & $2.218^{+0.001}_{-0.001}$ &  $-19.2^{+0.3}_{-0.3}$   &  $-1.6^{+0.2}_{-0.2}$   &  $42.81^{+0.07}_{-0.07}$   &  $180^{+210}_{-90}$  &  $230^{+20}_{-20}$   & NL-SF  \\
 \hline 
 SA22-NB5-14  & 22\,15\,22.60  & $+01\,31\,06.7$  & $3.114^{+0.001}_{-0.001}$ &  $-23.7^{+0.2}_{-0.2}$   &  $-2.4^{+0.1}_{-0.1}$   &  $44.21^{+0.06}_{-0.06}$   &  $230^{+40}_{-40}$  &  $4220^{+300}_{-290}$   & BL-AGN  \\
 SA22-NB5-18  & 22\,22\,02.72  & $-00\,27\,19.7$  & $3.096^{+0.001}_{-0.001}$ &  $-23.4^{+0.2}_{-0.2}$   &  $-2.5^{+0.1}_{-0.1}$   &  $44.03^{+0.06}_{-0.06}$   &  $110^{+20}_{-20}$  &  $1820^{+80}_{-80}$   & BL-AGN  \\
 SA22-NB5-10  & 22\,09\,19.01  & $-00\,06\,16.3$  & $3.102^{+0.001}_{-0.001}$ &  $-23.3^{+0.2}_{-0.2}$   &  $-1.1^{+0.1}_{-0.1}$   &  $43.99^{+0.07}_{-0.07}$   &  $90^{+16}_{-15}$  &  $1790^{+100}_{-90}$   & BL-AGN  \\
 BH-NB5-4(BR3)  & 14\,32\,30.55  & $+33\,39\,57.3$  & $3.086^{+0.001}_{-0.001}$ &  $-22.7^{+0.2}_{-0.2}$   &  $-2.4^{+0.5}_{-0.5}$   &  $43.78^{+0.06}_{-0.06}$   &  $150^{+30}_{-30}$  &  $690^{+80}_{-70}$   & NL-AGN  \\
 CALYMHA-383  & 02\,17\,32.39  & $-05\,12\,50.8$  & $3.222^{+0.004}_{-0.004}$ &  $-24.3^{+0.2}_{-0.2}$   &  $-1.7^{+0.1}_{-0.1}$   &  $43.54^{+0.06}_{-0.06}$   &  $26^{+4}_{-4}$  &  $4300^{+700}_{-700}$   & BL-AGN  \\
 BH-NB5-6  & 14\,33\,21.85  & $+33\,54\,20.3$  & $3.125^{+0.001}_{-0.001}$ &  $-23.2^{+0.2}_{-0.2}$   &  $0.7^{+0.5}_{-0.5}$   &  $43.56^{+0.06}_{-0.06}$   &  $37^{+6}_{-6}$  &  $2700^{+400}_{-400}$   & BL-AGN  \\
 BH-NB5-10  & 14\,33\,23.82  & $+33\,38\,47.2$  & $3.100^{+0.001}_{-0.001}$ &  $-23.2^{+0.2}_{-0.2}$   &  $-2.1^{+1.5}_{-1.5}$   &  $43.45^{+0.06}_{-0.06}$   &  $29^{+5}_{-5}$  &  $470^{+20}_{-10}$   & NL-AGN  \\
 BH-NB5-16  & 14\,31\,06.50  & $+34\,04\,23.8$  & $3.085^{+0.002}_{-0.002}$ &  $-21.7^{+0.2}_{-0.2}$   &  $-2.3^{+0.9}_{-0.9}$   &  $43.19^{+0.06}_{-0.06}$   &  $42^{+8}_{-8}$  &  $460^{+20}_{-20}$   & NL-AGN  \\
 BH-NB5-27  & 14\,30\,32.58  & $+33\,59\,22.1$  & $3.131^{+0.001}_{-0.001}$ &  $-21.2^{+0.2}_{-0.2}$   &  $-0.6^{+1.5}_{-1.5}$   &  $43.07^{+0.07}_{-0.07}$   &  $57^{+14}_{-13}$  &  $340^{+10}_{-10}$   & NL-AGN  \\
 GN-NB5-6712  & 12\,36\,07.98  & $+62\,23\,14.1$  & $3.148^{+0.001}_{-0.001}$ &  $-20.7^{+0.2}_{-0.2}$   &  $-2.1^{+0.7}_{-0.7}$   &  $43.10^{+0.06}_{-0.06}$   &  $380^{+300}_{-150}$  &  $400^{+110}_{-80}$   & NL-SF  \\
 BH-NB5-34  & 14\,31\,03.87  & $+33\,34\,46.2$  & $3.076^{+0.001}_{-0.001}$ &  $-20.9^{+0.2}_{-0.2}$   &  $-1.2^{+1.4}_{-1.5}$   &  $43.02^{+0.07}_{-0.07}$   &  $50^{+12}_{-11}$  &  $420^{+30}_{-20}$(2)   & NL-SF  \\
 BH-NB5-37  & 14\,33\,24.35  & $+33\,39\,38.5$  & $3.147^{+0.001}_{-0.001}$ &  $-20.3^{+0.3}_{-0.3}$   &  $-2.4^{+3.1}_{-3.1}$   &  $43.03^{+0.07}_{-0.07}$   &  $80^{+30}_{-20}$  &  $360^{+10}_{-10}$   & NL-SF  \\
 GN-NB5-3378  & 12\,37\,21.68  & $+62\,13\,50.2$  & $3.151^{+0.001}_{-0.001}$ &  $-20.4^{+0.3}_{-0.3}$   &  $-3.2^{+1.1}_{-1.2}$   &  $42.84^{+0.07}_{-0.07}$   &  $120^{+70}_{-40}$  &  $300^{+20}_{-20}$   & NL-AGN  \\
 GN-NB5-5878  & 12\,36\,19.47  & $+62\,15\,01.8$  & $3.129^{+0.001}_{-0.001}$ &  $-21.3^{+0.2}_{-0.2}$   &  $-1.7^{+0.5}_{-0.5}$   &  $42.69^{+0.08}_{-0.08}$   &  $40^{+10}_{-10}$  &  $390^{+20}_{-20}$   & NL-SF  \\
 \hline 
\end{tabular}
\label{zspec_Lya_props}
\end{table*}

\subsection{Final Ly$\alpha$ sample}\label{final_sample}

Out of the 23 targeted sources, we spectroscopically confirm 21 as LAEs\footnote{For one LAE, BH-NB392-55, we do not have coverage of the Ly$\alpha$ line itself, but rather confirm it through strong C{\sc iv} emission; see Figure \ref{2D_full_ISIS}.}, with seven sources at $z\approx2.2$ and 14 sources at $z\approx3.1$. Our final sample of bright LAEs ($\rm L_{\rm Ly\alpha}\gsim10^{42.7}$\,erg\,s$^{-1}$; $\rm L_{\rm Ly\alpha}\gsim L^*_{\rm Ly\alpha}$) is thus composed of 21 sources. We present the sample in Table \ref{zspec_Lya_props} and provide both the catalogue and the reduced 1D spectra on-line, with the published version of the paper.

%
%
\begin{figure*}
\includegraphics[width=17.5cm]{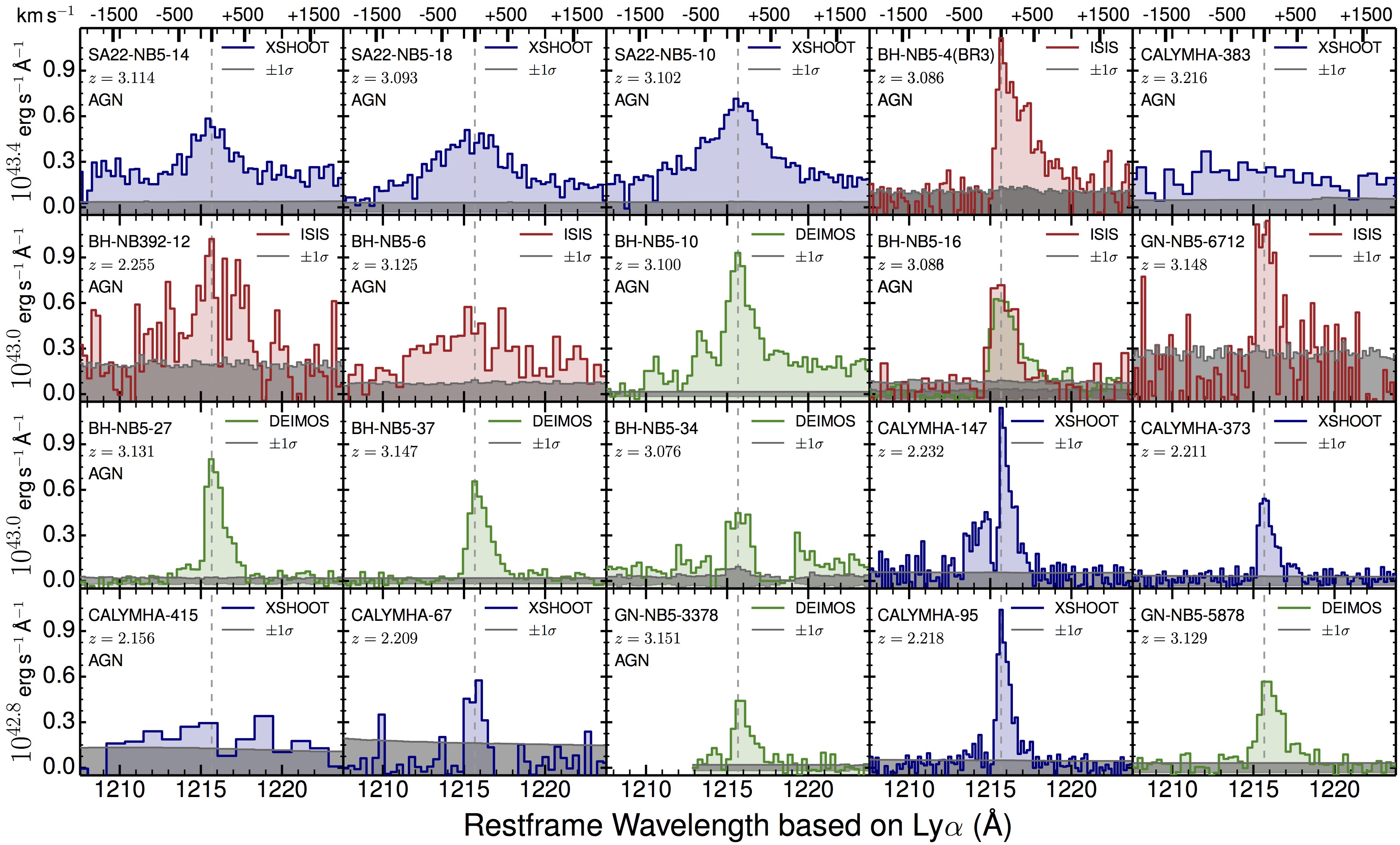}
\caption{Our 1D extracted, flux calibrated Ly$\alpha$ spectra for the three instruments (shown with different colours) used for our entire sample of 20 luminous LAEs with Ly$\alpha$ coverage. We order sources from the most to the least luminous in Ly$\alpha$ (based on the NB estimate prior to spectroscopy) from the top to bottom and left to right (note a decrease of 0.6\,dex from the upper to the lower panels), and we find significant differences in the line widths and profiles with decreasing luminosity. Overall, bright LAEs present a wide variety of Ly$\alpha$ line profiles, with FWHMs in excess of 1000\,km\,s$^{-1}$ at the highest Ly$\alpha$ luminosities, and narrow Ly$\alpha$ lines at lower Ly$\alpha$ luminosities. Five sources (25\%) show a blue-shifted Ly$\alpha$ component and CALYMHA-147 presents the highest blue-to-red ratio of $\approx0.6\pm0.1$. Note that BH-NB5-55 does not have spectral coverage at the Ly$\alpha$ wavelength and thus it is not shown. We show the spectra of BH-NB5-16 observed with both DEIMOS and ISIS, showing good agreement. Some sources with very low exposure time of just a few minutes (e.g. CALYMHA-415) yield relatively low S/N per spectral element, but their redshifts and other properties are securely determined by higher S/N detections of at least another line. Further properties of our Ly$\alpha$ sources are provided in Table \ref{zspec_Lya_props}.}
\label{1D_examples_fitting}
\end{figure*} 

\subsection{Line measurements and upper limits}

We explore our 1D spectra to fit single Gaussians and/or double Gaussians, depending on whether the emission is a doublet or not (Table \ref{table_lines}). When fitting a double Gaussian, we fix the doublet separation and set the FWHM to be the same for both lines. In order to avoid problems with OH sky lines and very low signal to noise (S/N) regions of the spectra, we mask regions with the strongest OH lines. We perturb each spectra in a window of $\pm4000$\,km\,s$^{-1}$ around each line being evaluated (see Table \ref{table_lines}) by independently varying each spectral element/data-point along their Gaussian probability (flux) and re-fit. We do this 10,000 times per source and per line and take the median, 16th and 84th percentiles as the best values and the lower and upper errors. We also compute the 0.6th and 99.4th percentiles, roughly corresponding to $-2.5$\,$\sigma$ to $+2.5$\,$\sigma$. Whenever the flux of the 0.6th percentile of a given line is consistent with zero or below we assign that line its 99.4th flux percentile as an upper limit. Table \ref{table_lines} provides the list of the lines that we use. As a further step, we visually inspect all lines and the best fits.

Not surprisingly, our most significant line is Ly$\alpha$, with a median S/N (integrated) of $\sim20$ (ranging from $\sim4$ to $\sim60$) across the entire sample and all instruments. The second most common line at high S/N is C{\sc iv} (see Table \ref{table_lines}) significantly detected in about half of the sample with a S/N from $\sim3$ to $\sim30$. Other high ionisation UV lines like N{\sc v}, He{\sc ii} and C{\sc iii}] are detected in $\sim4-6$ sources at typical S/N of $\sim3-10$ (BR3 yields S/N ratios of up to 50 in these lines). N{\sc iv} is only significantly detected with a S/N\,$\sim7$ in BH-NB5-4 (BR3), although tentatively detected at S/N$\sim2.5$ in another source; see Tables \ref{table_lines} and \ref{emission_lines_indiv}.

\subsection{UV luminosities and UV $\beta$ slope}

We compute the UV luminosity at rest-frame $\approx1500$\,{\AA} (M$_{\rm UV}$) for all our luminous LAEs. We use the magnitude in the closest observed band ($m$) to rest-frame 1500\,\AA, which approximately corresponds to $g$ for $z=2.2$ and $R$ for $z=3.1$ and compute:
\begin{equation}
{\rm M_{UV}}=m - [5\log_{10}(D_L[z])-5-2.5\log_{10}(1+z)]+f,  
\end{equation}
where $D_L$ is the luminosity distance in parsec, and $f$ is a correction factor from the fact that our filters may not trace 1500\,{\AA} exactly and have different widths. We use the results from \cite{Ilbert2009} and Ly$\alpha$ selected sources from \cite{Sobral2017_SC4K} to compare M$_{\rm UV}$ values computed using $g$ and $R$ and M$_{\rm UV}$ computed from the full SED fitting. We find averages of $f\approx0$ for $z\approx2.2$ ($g$) and $f\approx-0.2$ for $z\sim3.1$ ($R$). The dispersion on $f$ is $\approx0.2$. Errors on M$_{\rm UV}$ are computed by perturbing the appropriate $m$ and $f$ along their Gaussian distributions independently 10,000 times and calculating the 16th and 84th percentiles (see Table \ref{zspec_Lya_props}).

We estimate the rest-frame UV $\beta$ slope \citep[$f_{\lambda}\propto\lambda^{\beta}$; e.g.][]{Meurer1995,Meurer1997,Burgarella2005,Bouwens2009,Ono2010,Dunlop2012} with:
\begin{equation}
\beta=-\frac{m_1-m_2}{2.5\log_{10}(\lambda_{m_1}/\lambda_{m_2})}-2,
\end{equation}
by using $R$ ($m_1$) and $I$ ($m_2$) for $z\approx2.2$ and $I$ ($m_1$) and $z$ ($m_2$) for $z\approx3.1$ (observed $\lambda_{R}\approx6100$\,{\AA}, $\lambda_{I}=7600$\,{\AA}, $\lambda_{z}=9000$\,{\AA}, rest-frame $\approx1800,2200$\,{\AA} at each of the two redshift groups). We use these filters as they avoid strong contamination from C{\sc iv}, which would make strong C{\sc iv} emitters artificially blue if we obtain $\beta$ using rest-frame $\approx1500,2200$\,{\AA}. Errors on $\beta$ are computed with the same methodology as for M$_{\rm UV}$. Due to the slight differences in the filters and rest-frame, we add 0.1\,mag in quadrature to the final error (see Table \ref{zspec_Lya_props}).

\subsection{Estimating the Ly$\alpha$ escape fraction and $\rm \xi_{\rm ion}$}\label{Lya_escape}

We obtain a rough proxy for Ly$\alpha$ f$_{\rm esc,UV}$ based on M$_{\rm UV}$ (a more detailed investigation will be presented in Sobral et al. in prep., exploring H$\alpha$ and Balmer decrements). We obtain f$_{\rm esc,UV}$ by 1) converting M$_{\rm UV}$ to SFR (in M$_{\odot}$\,yr$^{-1}$) following \cite{Kennicutt1998} with a Salpeter IMF\footnote{Note that the choice of IMF is irrelevant due to step 2.}:
\begin{equation}
\rm SFR=(1.4\times10^{-28})(4\pi\times9.521\times10^{38})10^{-0.4(M_{UV}+48.6)}
\end{equation}
and by correcting for dust extinction; 2) converting Ly$\alpha$ to SFR assuming a 100\% escape fraction, case B and Salpeter IMF (see e.g. \citealt{Sobral2017_SC4K}), and 3) by obtaining the ratio between 2) and 1). For comparison, we also compute f$_{\rm esc}$ using the Ly$\alpha$ rest-frame EW as in \cite{Sobral2017} \citep[see also][]{Matthee2017c} and find excellent agreement.

Here we correct for dust extinction \citep[see discussions in][]{Matthee2016,An2017} based on our $\beta$ measurements. We use $A_{\rm UV}\approx4.43+1.99\beta$ \citep[][]{Meurer1999}, resulting in a bootstrapped $A_{\rm UV}\approx0.5^{+0.6}_{-0.2}$ (corresponding to the median $\beta=-2.0\pm0.2$\footnote{We exclude AGN and compute the bootstrapped median of $\beta$ as $-2.0\pm0.2$. If we use the entire sample we obtain a bootstrapped median of $\beta=-2.1\pm0.2$}, or $E(B-V)\approx0.05$; \citealt{Meurer1999}). This implies a correction to UV SFRs of our LAEs of $1.5^{+1.1}_{-0.3}$, and thus Ly$\alpha$ escape fractions (for our observed Ly$\alpha$ luminosities) lower by these factors compared to the no dust correction assumption.

Finally, we also estimate the ionisation efficiency ($\rm \xi_{\rm ion}$) of LAEs by following \cite{Matthee2017c}, using Ly$\alpha$ luminosity and the escape fraction to estimate the LyC luminosity and compare it to the dust corrected UV luminosity.


\section{Results} \label{results}

\subsection{AGN among the most luminous LAEs}

%

Some of the most luminous LAEs may be powered by AGN activity. We identify AGN within our luminous LAEs at $z\sim$\,2--3 as sources with i) broad Ly$\alpha$ lines (FWHM\,$>1000$\,km\,s$^{-1}$; e.g. Figure \ref{1D_examples_fitting} and Table \ref{zspec_Lya_props}), ii) X-ray detections or iii) emission line ratios associated with AGN activity, by exploring high ionisation lines (N{\sc v}, He{\sc ii}, C{\sc iv} and C{\sc iii}]); see Figure \ref{line_ratios_AGN_SF} \citep[see also e.g.][]{Feltre2016,Nakajima2017}.

\subsubsection{Broad-line and X-ray AGN}\label{BLAGN_class}

We start by identifying AGN using the broadness of the Ly$\alpha$ line (see further details in Appendix \ref{FWHMs_and_Voffsets}). We find that seven out of the 20 with Ly$\alpha$ FWHM measurements have FHWM clearly above 1000\,km\,s$^{-1}$ (see Figure \ref{1D_examples_fitting} and also Figures \ref{2D_spectra_WHTIS} and \ref{2D_spectra_DEI_XSHOOT} for 2D information), and thus we class them as broad-line AGN\footnote{Stellar winds may typically lead to de-projected gas velocities of $\sim100$ to $\sim1000$\,km\,s$^{-1}$, although in principle they can reach up to $\approx3000$\,km\,s$^{-1}$ \citep[][]{Heckman2003}. However, outflows are expected to lead to highly asymmetric lines, while the Ly$\alpha$ profiles of the broad-line LAEs are all very symmetric.} (BL-AGN) -- see Table \ref{zspec_Lya_props}. These sources have FWHMs ranging from $\sim1300$ to $\sim4000$\,km\,s$^{-1}$ (Figure \ref{1D_examples_fitting}). Our results imply a BL-AGN fraction of $35\pm11$\,\% within our full sample.

Our SA22 sources do not have X-ray coverage, as they are significantly away ($\sim1$\,deg in three different directions) from the SSA22 {\it Chandra} follow-up \citep[][]{Lehmer2009,Saez2015}, but they all show very broad Ly$\alpha$ emission (FWHM\,$\gsim1800$\,km\,s$^{-1}$), and thus are cleanly identified as AGN. Our GOODS-N sources have good {\it Chandra} coverage \citep[][]{Alexander2003}, but none of our three sources is detected in the 2\,Ms depth data. Our three sources in COSMOS are also not detected in deep X-ray observations \citep[see][]{Civano2016}, with the same holding for our three UDS sources, although for UDS only shallower {\it XMM} observations ($\approx3$ times shallower than COSMOS) are available \citep[][]{Ueda2008}. Finally, for our Bo\"{o}tes sources, we find one X-ray detection (BH-NB5-6, a broad-line AGN), while the other sources are not detected in the X-rays \citep[for a flux limit of $4\times10^{-15}$\,erg\,s$^{-1}$\,cm$^{-2}$, about 5 times shallower than COSMOS X-day data, see][]{Murray2005}. Therefore, using the available X-ray data, we find a single X-ray source (BH-NB5-6) out of 18 which have X-ray coverage, but we note that the available data are very heterogeneous, do not cover our most luminous LAEs in SA22, and are generally not deep enough to detect lower accretion and lower black mass AGN. Furthermore, the short duty cycles of X-rays can also lead to underestimating the AGN fraction using X-rays only \citep[][]{Shankar2009}.

\subsubsection{Narrow-line AGN}\label{NL_AGN_iden}

Our spectroscopy allows us to explore the nature of the sources further and search for AGN activity by using other emission lines besides Ly$\alpha$. We start by identifying sources with significant N{\sc v} (ionisation energy of 77.4\,eV) which, with observed ratios of $\rm N{\sc v}/Ly\alpha\gsim0.1$ we robustly class as AGN. This is because even the most extreme stellar populations fail to produce significant N{\sc v} emission ($\rm N{\sc v}/Ly\alpha<0.001$) due to the sharp cut-off of ionising photons at energies significantly below 77.4\,eV \citep[e.g.][]{Feltre2016}. This makes N{\sc v} one of the cleanest lines to identify AGN activity. We find that sources SA22-NB5-18, BH-NB5-6, $-$10, $-$16, $-27$ and CALYMHA$-$415 all show significant N{\sc v} emission (observed ratios $\rm N{\sc v}/Ly\alpha\approx0.1-0.6$). Half of these sources are BL-AGN (see Section \ref{BLAGN_class}), while BH-NB5-6 is also an X-ray source. We therefore find an AGN fraction of $32\pm11$\,\% based solely on high $\rm N{\sc v}/Ly\alpha$ ratios ($\rm N{\sc v}/Ly\alpha \gsim0.1$).

%
%
\begin{figure}
\includegraphics[width=8.33cm]{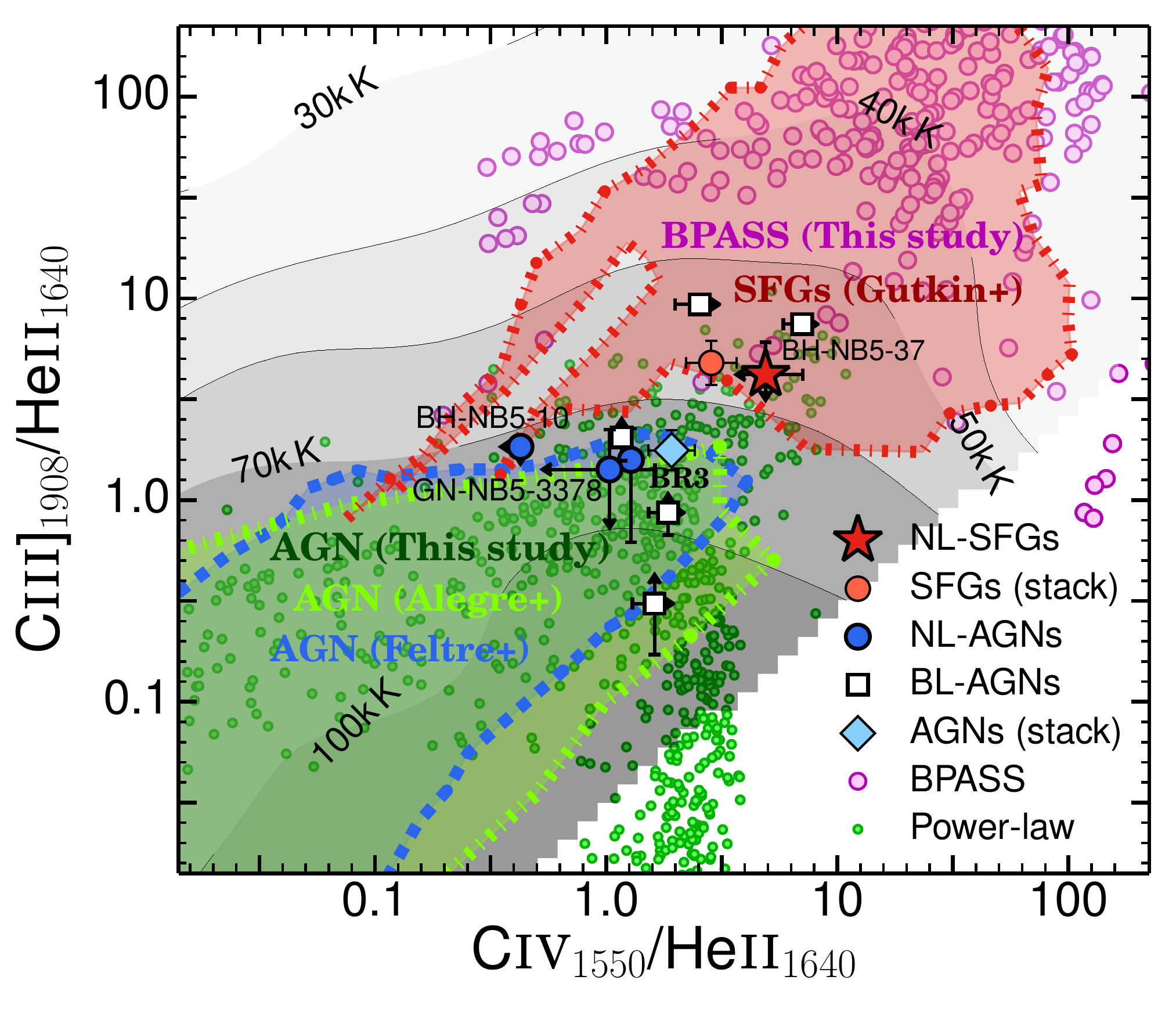}
\caption{Ratios between C{\sc iii}], C{\sc iv} and He{\sc ii} for the sources where we can constrain at least one of the lines in each of the ratios. We show the range of emission line ratios typical of power-law AGN \citep[][Alegre et al. in prep.]{Feltre2016} and those more typical of star-forming galaxies \citep[SFGs;][]{Gutkin2016}, together with an indication of the effective temperatures based on our black-body models. We also show the stack of all sources that we classify as SFGs, showing that they are within the region expected for SFGs and contrast them with the stack of all sources classified as AGN.}
\label{line_ratios_AGN_SF}
\end{figure} 

For the sources without N{\sc v} detections and with narrow Ly$\alpha$ lines ($<1000$\,km\,s$^{-1}$), we explore other UV rest-frame lines of lower ionisation energy. In Figure \ref{line_ratios_AGN_SF} we show emission line ratios of high ionisation UV lines which are in principle capable of further distinguishing between sources which are dominated by either AGN or star formation. We show four of our narrow-line emitters where we can measure/constrain both line ratios plotted. For comparison and interpretation, we also show the locations of power-law ionising sources (AGN) from \cite{Feltre2016}, along with the location for SFGs \citep[][]{Gutkin2016}. We also provide emission line ratios computed with {\sc cloudy} to sample both populations (see \S\ref{CLOUDY_lines_AGN_SF}). We show our luminous broad-line AGNs for which we can constrain the line ratios for comparison. Apart from BH-NB5-10, which was already classed as a NL-AGN due to the high N{\sc v}/Ly$\alpha$ ratio of $\approx0.17\pm0.03$, we identify two further narrow-line sources which are consistent with being NL-AGN: BR3, the brightest narrow-line LAE in our sample and GN-NB5-3378, with a much fainter Ly$\alpha$ luminosity. Both BH-NB5-10 and BR3 have emission-line FWHMs of $\approx600-700$\,km\,s$^{-1}$, higher than the other NL-emitters ($\approx100-300$\,km\,s$^{-1}$), also being brighter in UV and Ly$\alpha$. We find one source, BH-NB5-37, which is consistent with being dominated by star-formation, but suggests a relatively high effective temperature (see Figure \ref{line_ratios_AGN_SF}) and could still have an AGN source. For the other narrow-line sources without N{\sc v} where we do not detect high ionisation UV lines we classify them as SF as our observations should have detected high ionisation lines if they had a significant AGN contribution. Our stack of all these sources (see Figure \ref{line_ratios_AGN_SF}) suggests this is a reasonable assumption, as it places the overall sample of sources classified as SF in the SFG region in Figure \ref{line_ratios_AGN_SF}. We note, nonetheless, that even the stack of SFGs suggests high effective temperatures, and it is possible that even those LAEs have at least some AGN component.

\subsubsection{Final AGN and SF classifications}

Overall, we find seven broad line AGN and five narrow-line AGN, with a total of twelve AGN out of the sample of 20 classifiable sources. Our results therefore imply a total AGN fraction of $60\pm9$\,\% for luminous LAEs at $z\sim2-3$. Relatively high AGN fractions ($\sim20-50$\%) have also been found for bright LAEs at $z\sim0.3-1$ \cite[e.g.][]{Wold2014,Wold2017} and at $z\sim$\,2--3 \citep[e.g.][]{Ouchi2008,Konno2016,Sobral2017,Matthee2017b}, while for the much fainter, numerous LAEs the AGN fraction is typically no more than a few per cent \citep[e.g.][]{Wang2004,Nilsson2007,Ouchi2008,Oteo2015,Sobral2017_SC4K,Calhau2018}.

\subsection{The UV properties of luminous LAEs}\label{UV_relations}

\begin{figure}
\includegraphics[width=8.5cm]{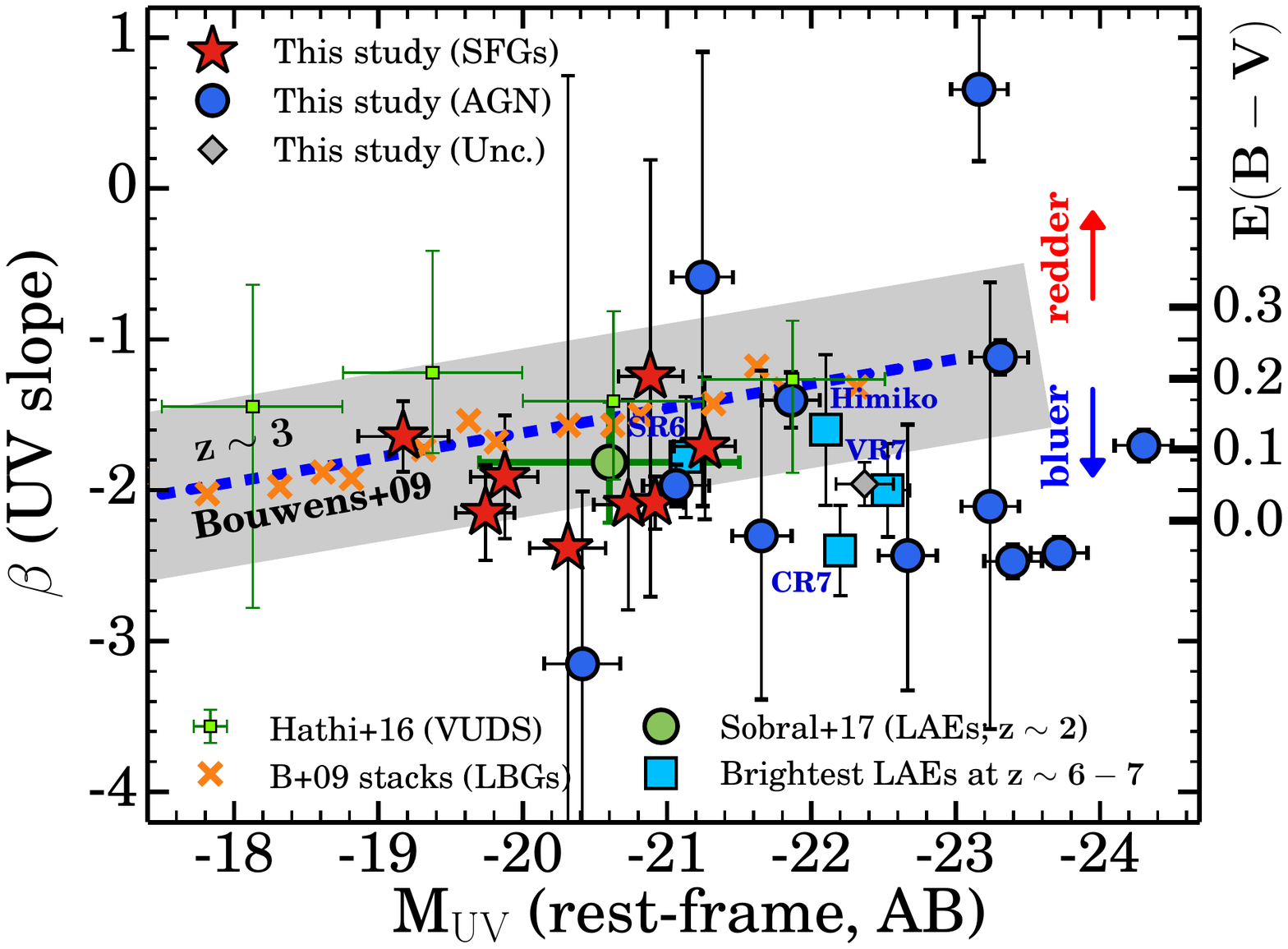}
\caption{The relation between rest-frame UV luminosities (M$_{\rm UV}$) and the UV $\beta$ slope for our luminous LAEs at $z\sim$\,2--3 and comparison with other studies: LBGs at $z\sim2-4$ \citep[][]{Bouwens2009}, LAEs at $z\sim2.2$ \citep[][]{Sobral2017}, LAEs within the UV continuum-selected VUDS sample at $z\sim2-2.5$ \citep[][]{Hathi2016_VUDS} and also the brightest $z=6-7$ LAEs \citep[][]{Matthee2017c}. We find that our sources are in general bluer than the population of Lyman-break selected sources at a similar redshift, with the deviation becoming stronger at the highest UV luminosities in the regime dominated by AGN, suggesting younger ages, low dust extinction and relatively unobscured AGN activity for the AGN. We find no significant relation between $\beta$ and $\rm M_{UV}$ for luminous LAEs with the slope of the potential relation being consistent with $0$ ($-0.07^{+0.11}_{-0.15}$).}
\label{UV_vs_Beta}
\end{figure}

On average, we find a bootstrapped median for the sample of SFGs of $\rm M_{UV}=-20.5^{+0.4}_{-0.3}$  and $\rm M_{UV}=-21.5\pm0.4$ for the full sample, with $\rm M_{UV}$ ranging from $-19.2$ to $-23.7$; see Table \ref{zspec_Lya_props} and Figure \ref{UV_vs_Beta}. We note that while luminous LAEs have Ly$\alpha$ luminosities of L$_{\rm Ly\alpha}^*$ or higher, they also have UV luminosities of roughly M$_{\rm UV}^*$ \citep[M$_{\rm UV}^*\approx-20.5$ at $z\sim2-3$, e.g.][]{Arnouts2005,Reddy2009,Parsa2016} and have a range of M$_{\rm UV}$ similar to the full sample of \cite{Sobral2017} and the sample of $\rm \gsim L_{\rm Ly\alpha}^*$ sources presented in \cite{Sobral2017_SC4K}, thus sharing properties with larger, photometric samples of LAEs. 

We find a bootstrapped median\footnote{The simple average for the full sample is $\beta=-1.9\pm0.7$} of $\beta=-2.0^{+0.3}_{-0.1}$ (the bluest source having $\beta\approx-3.2$ and the reddest having $\beta\approx0.7$; see Figure \ref{UV_vs_Beta}), revealing that our luminous LAEs are very blue on average, but with one outlier. The median $\beta$ that we find for luminous LAEs is in agreement with the full sample of LAEs at $z\sim2.2$ from \cite{Sobral2017}, and consistent with other studies focusing on lower luminosity Ly$\alpha$ selected sources \citep[e.g.][]{Venemans2005,Ouchi2008,Matthee2016}. Interpreting $\beta$ as a dust extinction indicator for the non-AGN sources suggests $E(B-V)=0.05^{+0.08}_{-0.05}$ \citep[][]{Meurer1999} or $A_{\rm UV}\approx0.5^{+0.6}_{-0.2}$ (see \S\ref{Lya_escape}), implying that the bulk of our luminous LAEs have little to no dust, and thus are potentially similar to fainter LAEs \citep[e.g.][]{Nilsson2007,Ono2010_lowz,Nakajima2012}. This is also consistent with Balmer decrement measurements/constraints for some of our sources (Matthee et al. in prep.). We list the individual M$_{\rm UV}$ and $\beta$ values for each source in Table \ref{zspec_Lya_props}.

In Figure \ref{UV_vs_Beta} we show the relation between M$_{\rm UV}$ and $\beta$ for luminous LAEs, and also compare our sources with the literature \citep[][]{Bouwens2009,Matthee2015,Matthee2017c,Hathi2016_VUDS,Sobral2015,Sobral2017}. We find little to no correlation between M$_{\rm UV}$ and $\beta$ for our luminous LAEs as a whole, with our sources being generally very blue, irrespective of UV luminosity (Figure \ref{UV_vs_Beta}). We further test this by fitting a linear relation to 10,000 realisations, each varying each individual pair of ($\rm M_{UV}$, $\beta$) randomly and independently within their full probability distributions. For the full sample (SFGs only) we find a slope of $-0.07^{+0.11}_{-0.15}$ ($-0.1^{+0.4}_{-0.2}$) for the relation between $\rm M_{UV}$ and $\beta$, fully consistent with no relation for either the full sample or just the sample of star-forming galaxies. We also find no significant relation when studying the AGNs only.

Our results are in very good agreement with \cite{Hathi2016_VUDS} who found a slope of $0.00\pm0.04$ for their global sample, consistent with our results of $-0.07^{+0.11}_{-0.15}$ for a much more luminous sample. Our luminous LAEs are significantly brighter in Ly$\alpha$ than essentially all LAEs in \cite{Hathi2016_VUDS} which come from a sample that is UV continuum-selected. We also note that we find luminous LAEs to be even bluer than sources studied in \cite{Hathi2016_VUDS}. While our results for SFGs are in reasonable agreement with those of \cite{Bouwens2009} for LBGs within the error bars (see Figure \ref{UV_vs_Beta}), LAEs tend to be below the relation found for LBGs. We also find that luminous LAEs deviate most from the \cite{Bouwens2009} relation at the brightest UV luminosities, but this may be a simple consequence of the powering nature for these sources (e.g. AGN).

\subsection{The relation between Ly$\alpha$ and UV at $\bf z\sim2-3$}\label{UV_Lya_relations}

\begin{figure*}
\includegraphics[width=15.24cm]{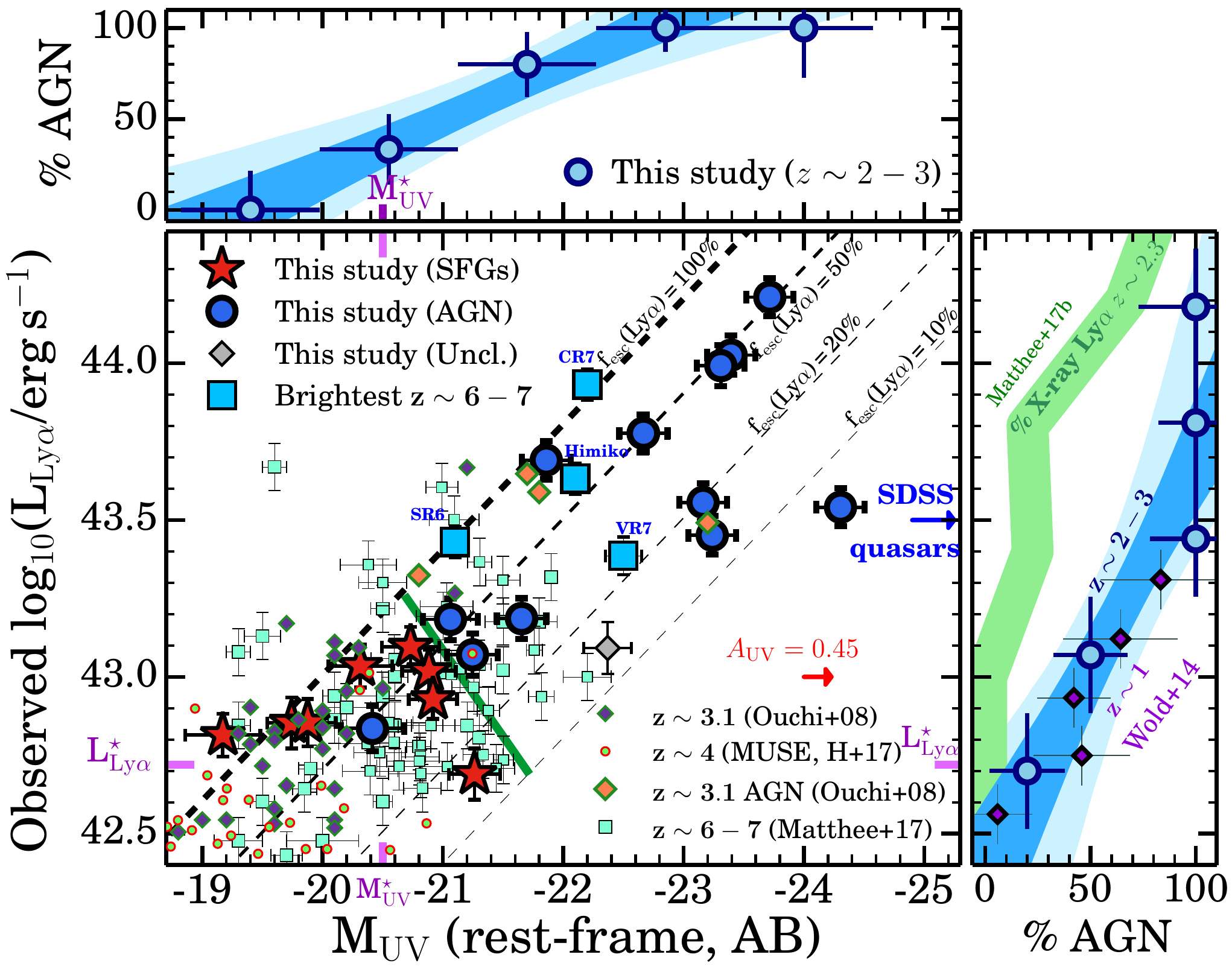}
\caption{The relation between rest-frame UV and Ly$\alpha$ luminosities. We find that Ly$\alpha$ roughly scales with UV for our sample at $z\sim$\,2--3, consistent with relatively high Ly$\alpha$ escape fractions, which we show (as dashed lines), from 10\% to 100\% assuming $\rm M_{\rm UV}$ traces star-formation and by correcting for dust with $\beta=-2.0$ (see arrow that shows how UV luminosity is increased by correcting for dust extinction and how the lines of constant Ly$\alpha$ escape fraction have shifted in the opposite direction). Our sources span a relatively unexplored region in the parameter space at $z\sim$\,2--3, in between the brightest LAEs found with MUSE \citep{Drake2017,Hashimoto2017} or the brightest typical LAEs \citep[e.g.][]{Ouchi2008,Trainor2015} and almost up to UV luminosities of the faintest quasars found at $z\sim$\,2--3 with e.g. SDSS \citep[e.g.][]{Richards2006}. Our AGN present the highest Ly$\alpha$ and UV luminosities within our sample; a simple dividing line ([-20.7,43.25],[-21.7,42.7]) is able to isolate all SFGs and all but one AGN from each other at $z\sim2-3$. The AGN fraction of luminous LAEs increases with both luminosities as $\rm f_{AGN}=(-0.30\pm0.07)({\rm M_{UV}+20.5})+(0.35\pm0.11)$ and  $\rm f_{AGN}=(0.78\pm0.22){(\rm \log_{10}(L_{Ly\alpha})-42.72)}+(0.24\pm0.14)$; we show the corresponding 1 and 2\,$\sigma$ contours. Qualitatively similar results have been found in recent studies \citep[see][]{Ouchi2008,Wold2014,Matthee2017b,Calhau2018}. We also compare our results with a literature compilation obtained by \citet{Matthee2017c} at higher redshift.}
\label{MUV_vs_Lya_Lum}
\end{figure*}

In Figure \ref{MUV_vs_Lya_Lum} we compare the Ly$\alpha$ luminosities of our sources with their rest-frame UV luminosities. Our sources span the relatively unexplored range between the more numerous LAEs, typically an order of magnitude fainter in Ly$\alpha$ and UV luminosities \citep[e.g.][]{Ouchi2008,Trainor2015,Drake2017,Hashimoto2017} and luminous quasars at $z\sim$\,2--3 \citep[][]{Richards2006}, which can have similar Ly$\alpha$ luminosities to our luminous LAEs \citep[e.g.][]{Borisova2016} but UV luminosities $\sim-25$ to $\sim-30$, and thus significantly brighter in the continuum.

The brightest LAEs in terms of M$_{\rm UV}$ within our sample are also the brightest in Ly$\alpha$, but the relation shows significant scatter, leading to a 2\,dex spread in M$_{\rm UV}$ for a $\approx1.7$\,dex spread in Ly$\alpha$ luminosity. Such scatter could be due to variations in dust content, the powering source of the ionising photons and/or due to different ionisation efficiencies \citep[see discussions in e.g.][]{Matthee2017,Sobral2017_SC4K}. One also expects that sources may have different Ly$\alpha$ escape fractions, even though it is known that LAEs (Ly$\alpha$-selected) tend to have typically high Ly$\alpha$ escape fractions \citep[][]{Nilsson2009,NILSSON2009LL,Wardlow2014,Trainor2015,Sobral2017}. We show indicative lines of constant Ly$\alpha$ escape fraction (correcting for dust extinction by using the UV $\beta$ slope; see Section \ref{Lya_escape}), which, under very simple assumptions, can easily explain the observed scatter. Many of our Ly$\alpha$ sources are consistent with having very high Ly$\alpha$ escape fractions which can be up to $\approx100$\%. The high escape fractions are likely an important factor in explaining why these sources are so Ly$\alpha$-bright. Alternatively, they may (also) have other powering sources hidden in the UV and/or a higher ionisation efficiency, $\xi_{\rm ion}$ \citep[e.g.][]{Matthee2017}. By estimating $\xi_{\rm ion}$ for the bright LAEs we find a bootstrapped\footnote{By correcting for dust extinction we obtain $\xi_{\rm ion}=10^{25.1\pm0.2}$\,Hz\,erg$^{-1}$.} $\xi_{\rm ion}=10^{25.4\pm0.2}$\,Hz\,erg$^{-1}$, with individual ionisation efficiencies for the star-forming LAEs varying from $\xi_{\rm ion}\approx10^{25.0}$\,Hz\,erg$^{-1}$ to $\xi_{\rm ion}\approx10^{25.9}$\,Hz\,erg$^{-1}$. These compare with values of $\xi_{\rm ion}=10^{25.0-25.3}$\,Hz\,erg$^{-1}$ for continuum selected LBGs at $z\sim2-5$ \citep[][]{Bouwens2015X, Shivaei2017,Nakajima2017}, $\xi_{\rm ion}=10^{24.8}$\,Hz\,erg$^{-1}$ for highly star-forming H$\alpha$ emitters at $z\approx2$ \citep[][]{Matthee2017} and $\xi_{\rm ion}\approx10^{25.5}$\,Hz\,erg$^{-1}$ for low luminosity LAEs at $z\sim3-6$ \citep[][]{Matthee2017,Harikane2017,Nakajima2018}.

%
%

%
%

Correcting for dust (see \S\ref{Lya_escape}) assuming $A_{\rm UV}=0.45$ ($A_{\rm UV}=0.0$) we find an average Ly$\alpha$ escape fraction of $50^{+20}_{-15}$\% ($70^{+30}_{-20}$\%) for our luminous SF LAEs at $z\sim$\,2--3. As an independent estimate, we also use the relation between Ly$\alpha$ escape fraction and the Ly$\alpha$ rest-frame EW presented in \cite{Sobral2017} which is independent of any assumptions regarding dust extinction, to find a bootstrapped Ly$\alpha$ escape fraction that is in perfect agreement ($50^{+20}_{-15}$\%)\footnote{We find a bootstrapped $\rm EW_0=90^{+44}_{-23}$\,{\AA}.}. These high escape fractions are comparable with those found for fainter LAEs. Some of the latest measurements include a Ly$\alpha$ escape fraction of $37\pm7$\% for a more general sample of fainter LAEs at $z\sim2$ \citep[][]{Sobral2017} and $\approx30$\% at $z\sim2-4$ \citep[e.g.][]{Wardlow2014,Kusakabe2015}. These results are much higher than the low escape fractions of $\approx2$\%\footnote{Directly measured with H$\alpha$.} for a more general population of star-forming galaxies selected through H$\alpha$ at $z\sim2$ \citep[][]{Matthee2016}, $\approx4-5$\% \citep[][]{Hayes2011,Sobral2017_SC4K} from a comparison of the evolution of the Ly$\alpha$ and UV luminosity functions at $z\sim2-3$, or $\approx5$\% for UV selected sources at $z\sim$\,2--3 \citep[e.g. VUDS;][]{Cassata2015}; see also \cite{An2017} and \cite{Sobral2017}. Our results thus show that luminous LAEs have Ly$\alpha$ escape fractions at least comparable or higher than the numerous low luminosity LAEs, hinting that at least the SF-dominated ones are likely ``scaled-up" or ``maximal" versions of the faint LAEs.

\subsection{Ly$\alpha$ and UV luminosities as AGN activity predictors for LAEs at $z\sim$\,2--3}\label{AGN_predictors}

Overall, our sample of luminous LAEs at $z\sim$\,2--3 reveals a high AGN fraction of $\approx60$\%. We now investigate how this fraction depends on both Ly$\alpha$ and UV luminosities. As Figure \ref{MUV_vs_Lya_Lum} shows, LAEs are essentially all AGN above a Ly$\alpha$ luminosity of $\approx10^{43.3}$\,erg\,s$^{-1}$ or for UV luminosities brighter than $\rm M_{UV}\approx-21.5$, qualitatively similar to results at different redshifts \citep[e.g.][]{Ouchi2008,Wold2014,Wold2017,Konno2016,Sobral2017,Matthee2017b}. These include both the narrow-line emitters, which are brighter in Ly$\alpha$ than in the UV, and BL-AGN, which are typically brighter in the UV. In Figure \ref{MUV_vs_Lya_Lum} we also show the rough location of the knee of the Ly$\alpha$ \citep[][]{Sobral2017_SC4K} and UV luminosity functions \citep[e.g.][]{Arnouts2005,Reddy2009,Parsa2016}. We find a rise of the AGN fraction from virtually 0 to 100\% happening just under 1\,dex in UV and Ly$\alpha$ luminosities, with the fraction below the knee of each luminosity function being $\approx0$\%. We quantify the relation between the AGN fraction and UV or Ly$\alpha$ luminosity in a similar way to other relations in this paper. We find that the AGN fraction among bright LAEs rises with M$_{\rm UV}$ significantly ($\sim4$\,$\sigma$) and is well parameterised (from $-19$ to $-23$) by $\rm f_{AGN}=(-0.30\pm0.07)({\rm M_{UV}+20.5})+(0.35\pm0.11)$. The relation between $\rm f_{AGN}$ and Ly$\alpha$ luminosity (significant at a $\sim3-4$\,$\sigma$) is well parameterised (from $\rm \log_{10}(L_{Ly\alpha})$ 42.5 to 44.0) by $\rm f_{AGN}=(0.78\pm0.22){(\rm \log_{10}(L_{Ly\alpha})-42.72)}+(0.24\pm0.14)$.

%
%
\begin{figure}
\includegraphics[width=8.3cm]{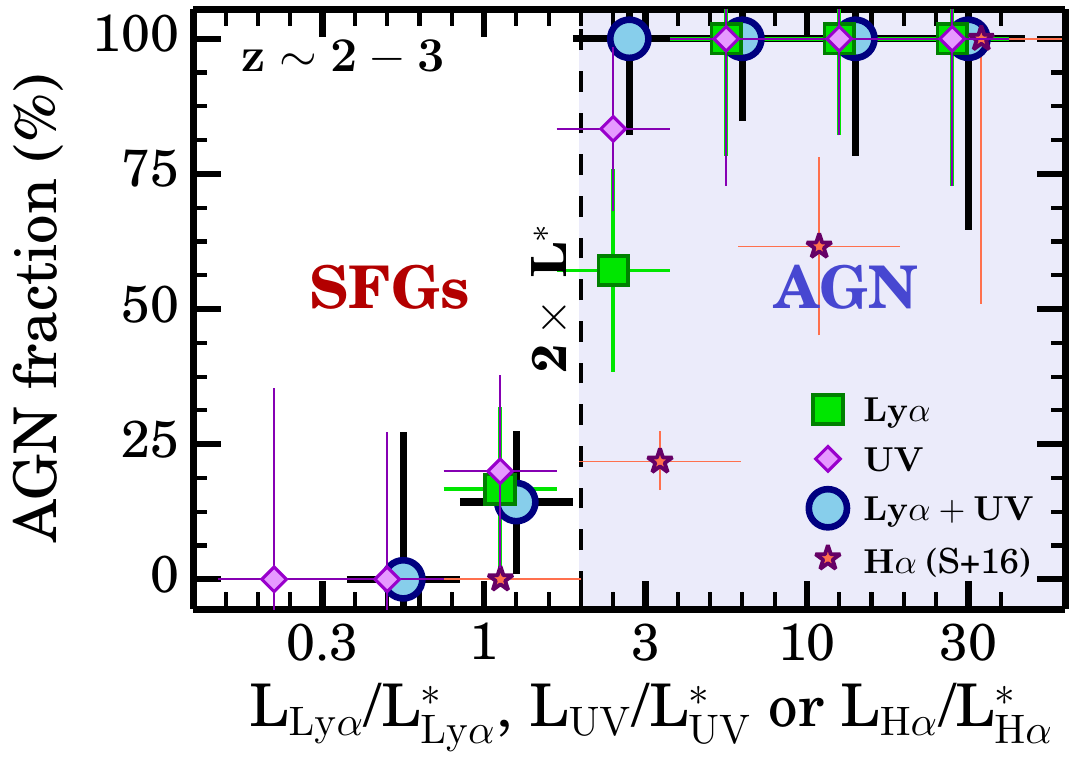}
\caption{The AGN fraction of luminous LAEs at $z\sim2-3$ as a function of the ratio between luminosity and the typical luminosity (L$^*$) for Ly$\alpha$, UV or the average of the two, and comparison to \citet{Sobral2016} for H$\alpha$ emitters with H$\alpha$ luminosity. We find that the transition from star-forming to AGN-dominated happens above twice the typical luminosity, with increasing Ly$\alpha$ luminosity resulting in a smoother growth than UV. When combining Ly$\alpha$ and UV we find a much sharper transition, well modelled by a step function. Our results provide a simple way to distinguish between AGN and SFGs and suggest an important physical transition at $\rm \approx2L^*$, likely linked with the maximal unobscured luminosity by a starburst at $z\sim2-3$ explaining the difference between LAEs and H$\alpha$ emitters.}
\label{AGN_FRACTIONS}
\end{figure} 

%
%
\begin{figure*}
\includegraphics[width=15.0cm]{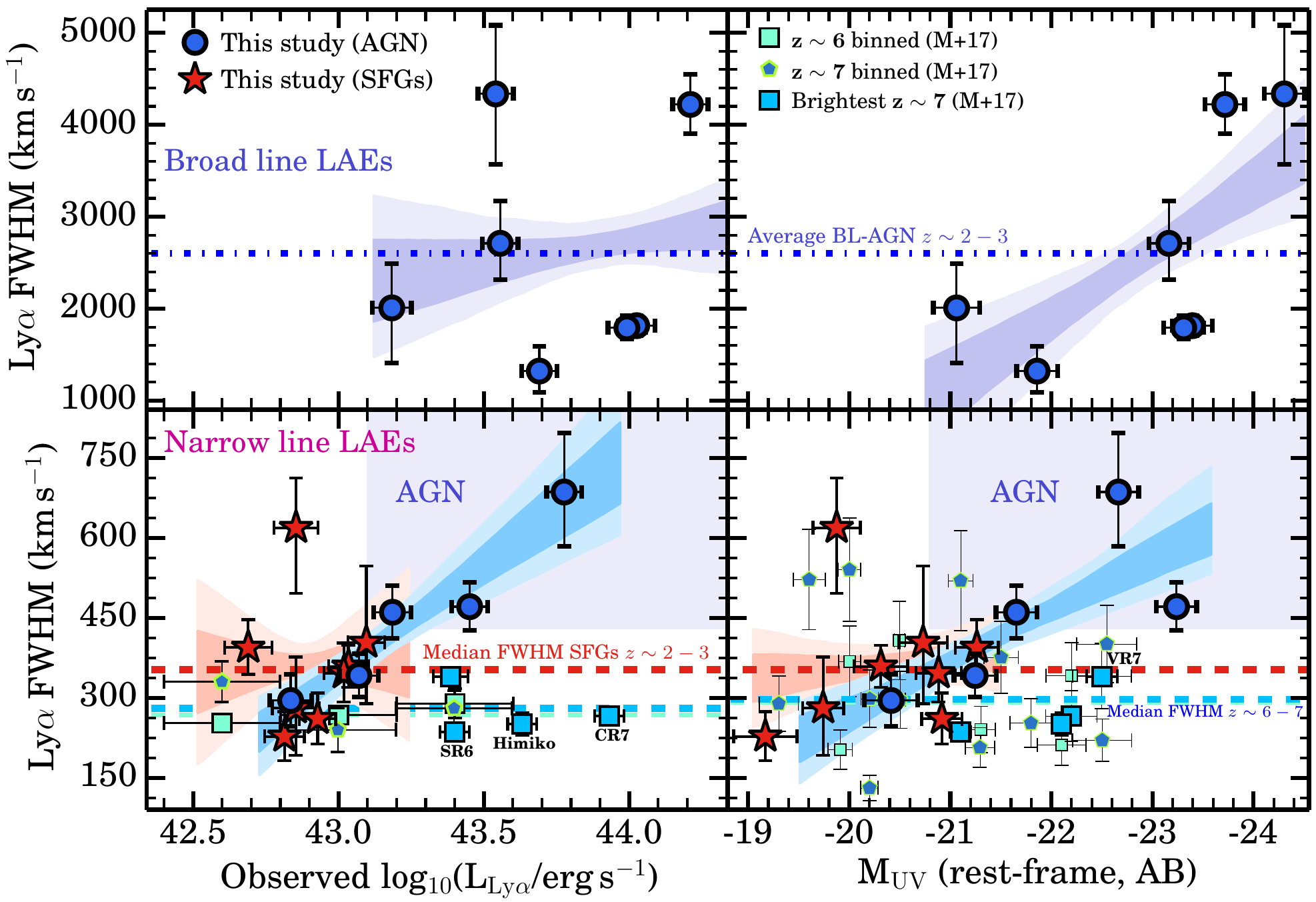}
\caption{Ly$\alpha$ FWHM versus Ly$\alpha$ and UV luminosities for $z=$\,2--3. On average, we find that Ly$\alpha$ FWHMs tend to increase with both Ly$\alpha$ and UV luminosities. However, we find that the trend is driven by AGN and that NL-AGN have the most statistically significant relations between luminosity and Ly$\alpha$ FWHM. We show the 2.3, 16, 84 and 97.7 percentiles from a Monte Carlo fitting routine. For BL-AGN, we find a weak to no correlation between FWHM and Ly$\alpha$ luminosity but a significant relation between FWHM and $\rm M_{UV}$. We find no significant relation between FWHM and any of the two luminosities for SFGs, although we note that a significant fraction of SFGs would be well fitted by an extrapolation of the relations found for NL-AGN. For comparison with sources selected with the same methods at higher redshift, we include a compilation presented in \citet{Matthee2017c} and also highlight the most luminous sources found at $z\sim6-7$, comparable in luminosities to some of our sources. Our results show a significant difference in the FWHMs of equally selected bright LAEs at fixed $\rm M_{UV}$ and $\rm L_{Ly\alpha}$ between $z\sim2-3$ and $z\sim6-7$.}
\label{MUV_andLLya_vs_LyaFWHM}
\end{figure*} 

We also compare our results with \cite{Wold2014,Wold2017} at $z\sim1$ and \cite{Matthee2017b} at $z\sim2.3$ (Figure \ref{MUV_vs_Lya_Lum}), which show very similar trends \citep[see also][]{Ouchi2008,Konno2016}. By comparing our results with $z\sim1$ (see Figure \ref{MUV_vs_Lya_Lum}) we find little to no evolution in the relation between the AGN fraction and the Ly$\alpha$ luminosity. This is apparently at odds with results from X-rays and radio \citep{Konno2016,Matthee2017b,Calhau2018}, which would hint for a strong evolution between $z\sim1$ and $z\sim2-3$. However, X-ray and radio selections are only able to recover a fraction of LAEs with AGN activity. Rest-frame UV spectra are crucial to cleanly identify sources with AGN signatures, as many do not show X-ray or radio emission down to even the deepest limits currently achieved over large fields. Our results further stress that X-ray identification of AGN \citep[e.g.][]{Konno2016,Sobral2017,Matthee2017b} must be taken as a lower limit to the AGN fraction, both due to short duty cycles and to the typically high accretion rates needed for detection in the X-rays at high redshift.

In order to further explore the relation between the AGN fraction and UV or Ly$\alpha$ luminosities, we present those together in Figure \ref{AGN_FRACTIONS}. We convert UV absolute magnitudes to luminosity and normalise by the corresponding L$^*_{\rm UV}$ ($\rm M^*_{\rm UV}=-20.5$, \citealt{Parsa2016}). We also normalise the Ly$\alpha$ luminosities by L$^*_{\rm Ly\alpha}=10^{42.72}$\,erg\,s$^{-1}$ \citep[][]{Sobral2017_SC4K}. The rise in the AGN fraction is found to be smoother with increasing Ly$\alpha$ luminosity than with UV luminosity (Figure \ref{AGN_FRACTIONS}), but the transition happens at the corresponding $\rm 2\times$\,L$^*$ in both cases and the AGN fraction starts rising above L$^*$. Using the average L/L$^*$ from combining UV and Ly$\alpha$ leads to the AGN fraction rising even sharply with increasing relative luminosity, from $11\pm10$\% below $\rm 2\times$\,L$^*$ to $100^{+0}_{-7}$\% above $\rm 2\times$\,L$^*$, suggesting a difference above 5\,$\sigma$. By perturbing each bin and re-fitting with i) a linear relation and ii) a step function with a transition at $\rm 2\times$\,L$^*$, we find that step functions result in reduced $\chi^2$ typically 2-3 times smaller than linear fits, with 92\% of all fits preferring a step-function. We note that combining the information on Ly$\alpha$ FWHMs with either Ly$\alpha$ or UV luminosities also leads to a sharper transition at $2\times$\,L$^*$ than when using a single luminosity to predict the AGN fraction.

Our results show that there is a relatively sharp transition in the nature of luminous LAEs at $2\times$\,L$^*$. Such transition suggests that either above L$^*$ AGN activity becomes more and more prominent \citep[see Figure \ref{AGN_FRACTIONS} and e.g.][]{Sobral2016}, and/or there is a relatively well defined physical limit in the output/observed luminosities from star-forming galaxies at $z\sim2-3$ which does not apply to AGN. This limit would likely be linked with i) the intrinsic maximal starburst before radiation pressure halts further activity \citep[see][]{Crocker2018} and perhaps even more importantly with ii) the maximal ``observable" or unobscured starburst which requires the highest possible output of LyC and UV photons without significant dust obscuration, so it can be seen as bright as possible in Ly$\alpha$ and UV. Thus, while there are star-forming galaxies that intrinsically produce much higher UV and Ly$\alpha$ luminosities than the observed $2\times$\,L$^*$ in the UV \citep[][]{Casey2014_REV}, a significant increase in SFR is linked with a similar or even higher increase of dust obscuration \citep[e.g.][]{Hopkins2001,GarnBest2010,Sobral2012}, leading to an observed UV and Ly$\alpha$ luminosity that can even decrease, thus limiting the maximum observed luminosities \citep[see e.g.][]{Nilsson2011}. If this is the case, the only sources that can still be observable as LAEs brighter than $1-2\times$\,L$^*$ require AGN activity. In these cases, due to the physics of the accretion disk or powerful outflows clearing out material, the AGN activity can lead to even larger observed UV and Ly$\alpha$ luminosities. Results from H$\alpha$ (non-resonant and less affected by dust) selected sources \citep[][]{Sobral2016} show an increase of the AGN fraction with increasing L$_{\rm H\alpha}$/L$^*_{\rm H\alpha}$ above $2\times$\,L$^*_{\rm H\alpha}$ at $z\sim2$ (Figure \ref{AGN_FRACTIONS}), but such increase is significantly slower and the AGN fraction only reaches 100\% by $20\times$\,L$^*_{\rm H\alpha}$, a factor 10 larger than seen for LAEs. We discuss these results further and implications for a physical limit for the UV and Ly$\alpha$ starburst observed luminosities in \S\ref{disc_AGN_SF}.

\subsection{The relation between Ly$\alpha$ FWHMs, UV and Ly$\alpha$ luminosities at $\bf z\sim2-3$}\label{FWHMs_vs_UV_Lya}

In Figure \ref{MUV_andLLya_vs_LyaFWHM} we investigate how the Ly$\alpha$ FWHM depends on Ly$\alpha$ and rest-frame UV luminosities for bright LAEs at $z\sim$\,2--3. We start by evaluating a potential simple linear relation between FWHM and each luminosity. We find a slope that is significantly ($>5$\,$\sigma$) away from zero (no-relation), with no realisation ($p<10^{-4}$) consistent with a zero slope. However, the strong trend could be a consequence of a sharp change in the population from narrow to broad-line emitters from low to high luminosities. In order to further investigate the trends we split the sample in i) SFGs, ii) NL-AGN and iii) BL-AGN and repeat the measurements/analysis. Our results show that there is no significant correlation between Ly$\alpha$ FWHM and either Ly$\alpha$ or UV luminosity for SFGs alone (see Figure \ref{MUV_andLLya_vs_LyaFWHM}), with the best-fit slope being fully consistent with zero within the 16 and 84 percentiles. For NL-AGN we find a strong correlation between FWHM and both Ly$\alpha$ and UV luminosities (see Figure \ref{MUV_andLLya_vs_LyaFWHM}), with the median slopes of all fits being $\approx4-5$\,$\sigma$ away from zero and with 100\% of realisations resulting in a positive slope. This relation may be related with outflows becoming more prominent and ejecting material at higher velocities for higher luminosity NL-AGN. Alternatively, high velocity outflows may lead to higher Ly$\alpha$ escape fractions, thus explaining the tighter correlation between Ly$\alpha$ luminosity and FWHM than UV luminosity with FWHM. It is worth noting, nonetheless, that there is a significant fraction of SFG LAEs that are well fitted by the range of fits performed to the NL-AGN, with many of these SFGs lying well within the 1\,$\sigma$ range of best fits to the NL-AGN. This could imply that the physics behind the correlation is similar in both NL-SFGs and NL-AGN, and could be driven by low velocity SF-driven outflows which may be roughly proportional to the LyC production (SFRs).

Our results for BL-AGN reveal no significant correlation between Ly$\alpha$ FWHM and Ly$\alpha$ luminosity, with $>30$\% of realisations being consistent with a zero or even an anti-correlation, while we find a relatively significant correlation between FWHM and $\rm M_{\rm UV}$ for BL-AGN, with the slope being $\sim3$\,$\sigma$ away from zero. For BL-AGN, our results may be explained by the physics of the accretion disk and the mass of the super-massive black holes: more massive black holes will produce higher FWHMs and higher UV luminosities, but the observed Ly$\alpha$ output may both be variable and will depend on orientation angle and the likely complex radiation transfer. This could explain why the FWHM correlates well with UV luminosity, but little with observed Ly$\alpha$.

We also compare our results at $z\sim$\,2--3 with the literature at higher redshift for LAEs with comparable Ly$\alpha$ and UV luminosities (Figure \ref{MUV_andLLya_vs_LyaFWHM}) and selected in the same way as our sources \citep[][]{Matthee2017c}. We find that at $z\approx5.7-6.6$ there is little to no relation between FWHM and UV or Ly$\alpha$ luminosity. Furthermore, we find an offset between the median FWHM at $z\sim$\,2--3 and $z\sim6-7$. Such an offset is particularly striking at the highest Ly$\alpha$ luminosities ($>10^{43.3}$\,erg\,s$^{-1}$) and the highest UV luminosities ($<-21.5$), as the FWHMs at $z\sim6-7$ are relatively narrow ($\sim200-300$\,km\,s$^{-1}$; \citealt{Matthee2017c}), while the median at $z\sim$\,2--3 is above 1000\,km\,s$^{-1}$. This suggests that the powering sources (and the ISM/IGM conditions) needed to result in exceptionally high Ly$\alpha$ and UV luminosities are different at $z\sim6-7$ and $z\sim$\,2--3, and that high Ly$\alpha$ and UV luminosities can be achieved at high redshift without potentially very energetic/high velocity outflows (see \S\ref{AGN_SF_evolution}). These may be connected with e.g. higher $\xi_{\rm ion}$, higher Ly$\alpha$ escape fractions \citep[see discussions in][]{Sobral2017_SC4K} and potentially lower metallicity environments which ultimately may result in higher luminosities for unobscured starbursts at high redshift (see \S\ref{AGN_SF_evolution} for further discussion).


\section{The nature of the brightest LAEs with rest-frame UV lines} \label{UV_lines_nat}

\subsection{Velocity offsets of luminous LAEs at $\bf z\sim$\,2--3}

We use non-resonant lines (when detected) to derive the systemic redshift of each source, and then use them to calculate the shift of the peak of Ly$\alpha$ from each line (see Appendix \ref{FWHMs_and_Voffsets}). We find such shifts to be generally positive, i.e., for Ly$\alpha$ to generally be redshifted compared to the other emission lines \citep[e.g.][]{Trainor2016}, but we find some notable exceptions, particularly for sources where C{\sc iv} and/or He{\sc ii} are redshifted more than Ly$\alpha$ by $\sim100$\,km\,s$^{-1}$ (e.g. GN-NB5-3378). In general, we find C{\sc iv} to be redshifted by $\approx100$\,km\,s$^{-1}$ compared to other lines which should be tracing the systemic redshift better. This is not surprising since C{\sc iv} is also a resonant line, and thus has a behaviour similar to Ly$\alpha$.

For the narrow LAEs at $z\sim2.2$, we find low velocity offsets between Ly$\alpha$ and other rest-frame UV lines from $\approx30$\,km\,s$^{-1}$ to $\approx250$\,km\,s$^{-1}$ with an average of $140^{+80}_{-30}$\,km\,s$^{-1}$; these are sources with the most symmetric and narrowest Ly$\alpha$ FWHM ($\approx200-300$\,km\,s$^{-1}$) and also include sources with significant blue wings \citep[see also e.g.][]{Verhamme2017}. For the sample at $z\sim3$ we find a larger variety of velocity offsets, ranging from $\approx-250$\,km\,s$^{-1}$ to $\approx800$\,km\,s$^{-1}$ and an average of $\approx100^{+300}_{-240}$\,km\,s$^{-1}$. The higher incidence of blue-shifted Ly$\alpha$ in emission for some of the $z\sim3$ sources might be linked with their higher luminosities and potentially higher velocity outflows.

Overall, we find typical Ly$\alpha$ to systemic velocity offsets of $\approx120$\,km\,s$^{-1}$ but with a number of offsets (for seven LAEs) consistent with zero, implying that Ly$\alpha$ is likely escaping close to systemic. The velocity offsets we obtain can be compared with typical velocity offsets of $\approx450\pm50$\,km\,s$^{-1}$ from UV selected galaxies at similar redshifts \citep[e.g.][]{Erb2010,Steidel2010} and $\sim200$\,km\,s$^{-1}$ for much fainter Ly$\alpha$ selected sources \citep[see][]{Trainor2015,Guaita2017,Verhamme2017}. We conclude that luminous LAEs have Ly$\alpha$ velocity offsets in respect to systemic which are similar, or even smaller, to those of the faint Ly$\alpha$ selected sources, and much lower than that of the more general Lyman-break population at similar UV luminosities. Such low velocity offsets may be a consequence of the relatively high Ly$\alpha$ escape fractions \citep[see e.g.][]{Behrens2014,Verhamme2015,Rivera-Thorsen2017}, which in turn provide conditions for the high observed Ly$\alpha$ luminosities.

%
%
\begin{figure}
\includegraphics[width=8.35cm]{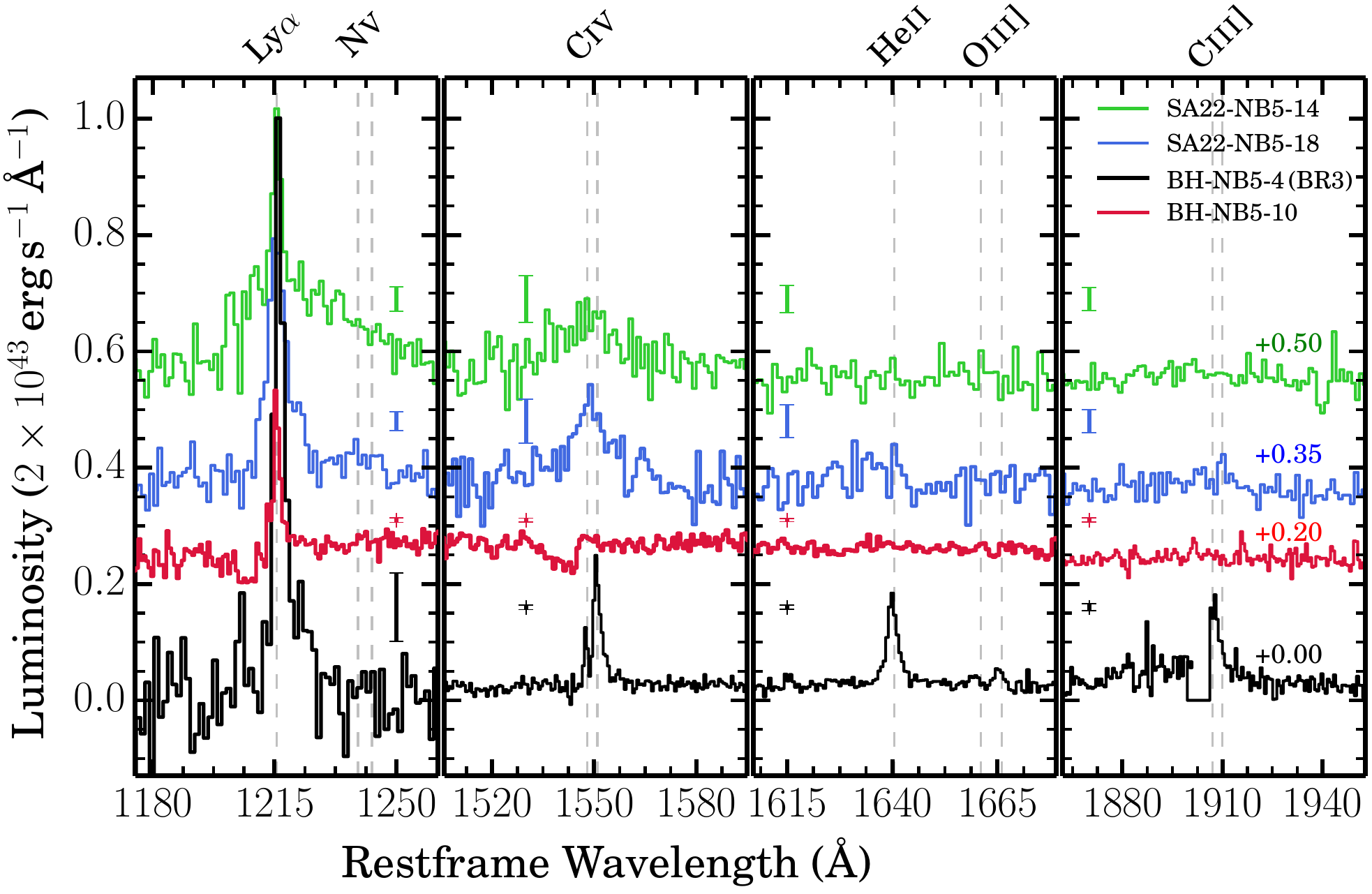}
\caption{Comparison between some of our brightest LAEs at $z=3.1$ (all classified as AGN), showing a diversity of FWHMs, continuum and also other rest-frame UV lines. We show the 1\,$\sigma$ error on the 1D for each rest-frame wavelength window so it is easier to distinguish between noise and real emission and absorption features. Note that spectra have been shifted up in order to avoid overlap. The brightest broad LAE (SA22-NB5-14) shows significant transmission blue-ward of Ly$\alpha$, while the second brightest (SA22-NB5-18) shows much less transmission, and the narrow line emitters show little to no transmission just blue-ward of Ly$\alpha$.}
\label{BR3_and_friends}
\end{figure} 

\subsection{The nature of the brightest LAEs with individual emission-line detections}\label{CLOUDY_lines_AGN_SF}

In Figure \ref{BR3_and_friends} we show our most luminous narrow-line LAE (BR3), which shows strong C{\sc iv}, He{\sc ii}, C{\sc iii}] and weaker O{\sc iii}], revealing a highly ionising powering source. No N{\sc v} is detected and all lines have FWHMs similar to Ly$\alpha$, of $\approx600$\,km\,s$^{-1}$. We also show even brighter broad-line emitters, and one of the brightest in the UV (BH-NB5-10), which shows large blue-shifted absorption features at C{\sc iv} and Ly$\alpha$. Figure \ref{BR3_and_friends} reveals the large variety in the rest-frame UV spectra of the most luminous LAEs at $z\sim$\,2--3, which we find to be consistent with being powered by AGN. 

In order to interpret the spectra, we conduct photo-ionisation modelling (see also \citealt{Nakajima2017}) using i) power-law spectra (to model AGN), ii) {\sc bpass} \citep[][]{Eldridge2009,Eldridge2012,Stanway2016} stellar models, including binary evolution and iii) black-bodies with a range of effective temperatures. We explore a relatively wide range of physical conditions by using the {\sc cloudy} (v 13.05) photo-ionisation code \citep{CLOUDY1998,CLOUDY2013}. Further details are given in Alegre et al. (in prep.) and in Appendix \ref{CLOUDY_AP}. We summarise the physical conditions  we explore in Table \ref{CLOUDY_table} \citep[see also][]{Feltre2016,Gutkin2016,Nakajima2017}.

We estimate physical conditions within bright LAEs by using the line ratios presented in Table \ref{emission_lines_indiv}. In addition, we use upper limits to provide either upper or lower limits on line ratios. We then explore the multi-dimensional line ratio parameter space to find all {\sc cloudy} models that produced line ratios consistent with observations within 3\,$\sigma$ of each line ratio, upper or lower limits of the line ratios. We then compute the best physical parameter (see Table \ref{CLOUDY_table}) and typical uncertainties by taking the median, 16 and 84 percentiles of each physical parameter within all models that result in line ratios within 3\,$\sigma$ of the observations. This means that our approach is statistically more robust than simply trying to find the ``best-fit'' of {\sc cloudy} outputs to observations and more conservative in uncertainties than focusing on e.g. a single combination of line ratios. We also compute the fraction of emission line ratios that each type of model is able to reproduce within 1$\sigma$ of the constrained line ratios; we use these to discuss which model best described each source or stack of sources.

BR3 is best described by hot black-body models or power-law models. BPASS photoionisation models fail to reproduce 40\% of the emission line ratios available, re-enforcing BR3's AGN nature. Power-law models suggest BR3 has a power-law slope of $\alpha=-1.2^{+0.2}_{-0.3}$, an ionisation parameter of $\log\,U=-2.2^{+0.2}_{-0.1}$ and a gas metallicity of $\rm \log(O/H)+12=8.5\pm0.2$. Black-body models suggest $\rm T_{eff}=135\pm10$\,kK, an ionisation parameter of $\log\,U=-2.0^{+1.3}_{-0.3}$ and a gas metallicity of $\rm \log(O/H)+12=8.3^{+0.4}_{-0.1}$. The results suggest that BR3 is a highly ionising source with a high effective temperature, while its gas is consistent with being mildly sub-solar at roughly 0.5\,$Z_{\odot}$.

%
%
\begin{table}
 \centering
  \caption{Parameters and ranges used for our photo-ionisation {\sc cloudy} \citep{CLOUDY1998,CLOUDY2013} modelling. We vary density, metallicity and the ionisation parameter ($\log\,U$) for both the star-like ionisation (here modelled with BPASS, but also with black bodies of varying temperature from 20 to 160K) and for an AGN-like ionisation (here modelled with with a power-law slope).}
  \begin{tabular}{cc}
  \hline
   \bf Parameter & \bf Range used for all models   \\
 \hline
   \noalign{\smallskip}
  Density (n$_{\rm H}$\,cm$^{-3}$) & 100, 300, 1000  \\
  Metallicity ($\rm \log\,Z_{\odot}$)  &  $-$2 to $+$0.5 (steps of 0.05)  \\ 
  Ion. param. ($\log\,U$) & $-$5 to $+$2 (steps of 0.2) \\
 \hline
  \bf Type of model & \bf Range used   \\
  \hline
   Black body (Temp., K) & 20k to 160k (steps of 1k) \\
   Power-law (slope) & $-$2.0 to $-$1.0 (steps of 0.05)  \\
   BPASS ($\log$\,Age, yr) & 6.0 to 9.0 (steps of 0.1) \\
  \hline 
\end{tabular}
\label{CLOUDY_table}
\end{table}

%
%
\begin{figure*}
\includegraphics[width=17.2cm]{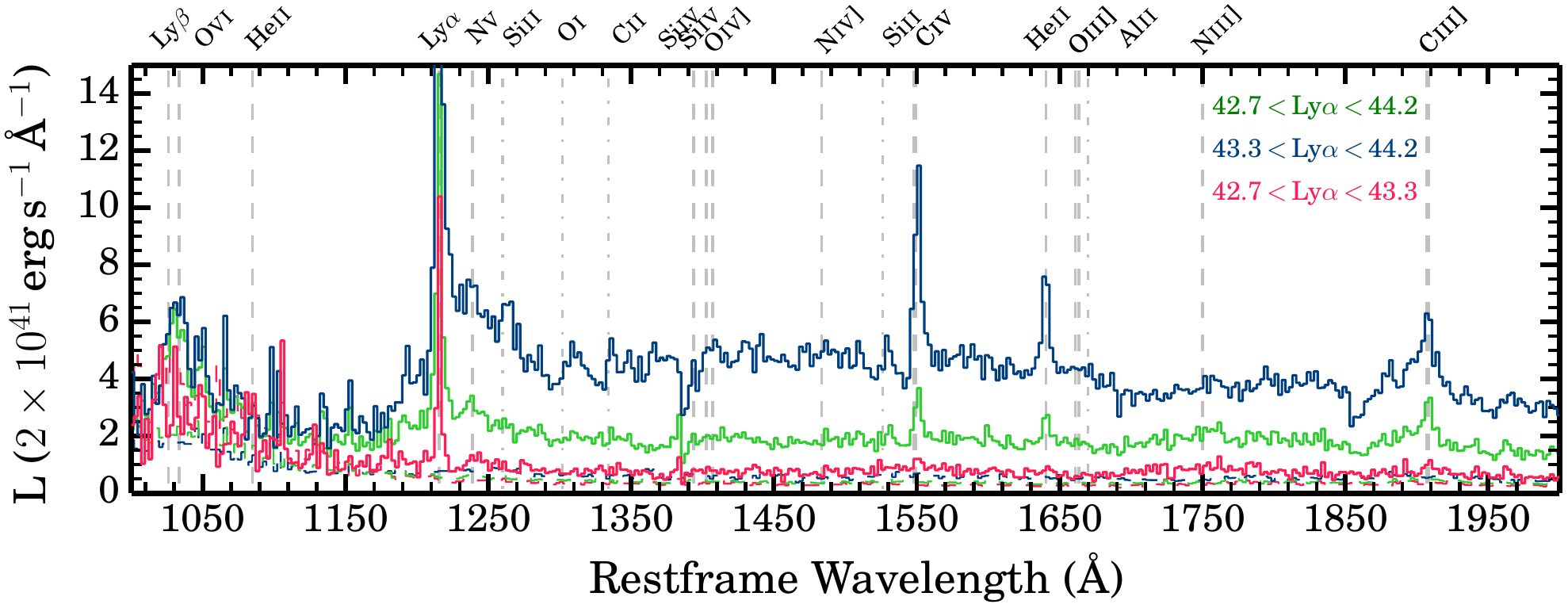}
\caption{Examples of weighted (by the inverse of the noise) average stacks by splitting our sample of $z\sim$\,2--3 luminous LAEs in two based on the average Ly$\alpha$ luminosity and also by stacking the full sample. We indicate the rest-frame wavelengths of the main rest-frame UV nebular and ISM lines with dashed and dot-dashed lines, respectively. The horizontal dashed lines show the 1\,$\sigma$ error per resolution element of each stack. We find a strong increase in the high ionisation UV emission lines with increasing Ly$\alpha$ luminosity, while blue-shifted ISM absorption lines are also clearer and more blueshifted for the stack of the most luminous LAEs.}
\label{1D_stack_all}
\end{figure*} 

%
%
\begin{table*}
 \centering
  \caption{Results from the stacking of our sources in different sub-samples and using all constrained UV lines (see Table \ref{emission_lines_indiv}) in order to extract likely physical conditions. We provide both line ratios for the stacks and potential physical conditions from our {\sc cloudy} modelling; see Section \ref{CLOUDY_lines_AGN_SF} and Table \ref{CLOUDY_table}. Effective temperatures (T$_{\rm eff}$) are indicative, coming from the warm ionised inter-stellar medium and from our black-body ionising sources. Gas metallicities are given in $\log$\,(O/H)+12 with solar being 8.7.}
  \begin{tabular}{@{}cccccccccc@{}}
  \hline
 Stack & N{\sc v}/Ly$\alpha$  & C{\sc iv}/He{\sc ii}  & C{\sc iii}]/He{\sc ii}   & $\log\,U$ & Gas Metallicity  & Burst Age & Power-law  & T$_{\rm eff}$ \\  
      &   &     &   &   &   & (Myrs) & $\alpha$ & (kK)  \\
      \hline
   Full sample  &     $0.11\pm0.01$  &     $1.7\pm0.2$  &     $5.2\pm0.4$  &    $-0.7^{+0.5}_{-0.4}$  &  $8.7^{+0.3}_{-0.1}$  & $6^{+25}_{-4}$  & $-1.4^{+0.3}_{-0.3}$  & $150^{+10}_{-50}$  \\ 
      \hline
     All SFGs  &     $<0.16$  &     $2.9\pm0.8$  &     $4.8\pm1.4$  &    $-3.0^{+1.6}_{-0.9}$  &  $8.2^{+0.5}_{-0.3}$  & $20^{+40}_{-15}$  & ---  & ---  \\ 
  All AGNs  &     $0.11\pm0.02$  &     $2.3\pm0.2$  &     $3.0\pm0.2$  &    $0.6^{+0.5}_{-0.5}$  &  $8.8^{+0.1}_{-0.1}$  & ---  & $-1.4^{+0.4}_{-0.2}$  & $70^{+70}_{-10}$  \\ 
   \hline
   $42.6<\rm Ly\alpha<43.3$  &     $0.05\pm0.02$  &     $2.5\pm0.9$  &     $2.5\pm0.8$  &    $-1.3^{+0.1}_{-0.4}$  &  $7.5^{+0.5}_{-0.1}$  & $4^{+2}_{-2}$  & $-1.7^{+0.3}_{-0.1}$  & $130^{+20}_{-20}$  \\ 
   $43.3<\rm Ly\alpha<44.2$  &     $0.22\pm0.02$  &     $2.0\pm0.1$  &     $2.6\pm0.2$  &    $-0.6^{+0.1}_{-0.1}$  &  $8.7^{+0.1}_{-0.1}$  & $3^{+1}_{-2}$  & $-1.5^{+0.3}_{-0.3}$  & $155^{+5}_{-5}$  \\ 
        \hline
   $-21.5<\rm M_{UV}<-19.1$  &     $<0.12$  &     $<1.9$  &     $2.1\pm0.7$  &    $-2.0^{+1.3}_{-0.6}$  &  $8.7^{+0.1}_{-0.5}$  & $60^{+250}_{-55}$  & $-1.5^{+0.3}_{-0.3}$  & $120^{+20}_{-20}$  \\ 
   $-24.4<\rm M_{UV}<-21.5$  &     $<0.18$  &     $2.2\pm0.2$  &     $3.1\pm0.3$  &    $-1.1^{+0.4}_{-1.1}$  &  $8.3^{+0.2}_{-0.1}$  & $13^{+1}_{-2}$  & $-1.0^{+0.1}_{-0.6}$  & $90^{+50}_{-20}$  \\ 
        \hline
   $200<\rm FWHM<1000$  &     $0.10\pm0.01$  &     $<0.70$  &     $7.2\pm1.7$  &    $-0.0^{+0.2}_{-0.2}$  &  $8.9^{+0.1}_{-0.1}$  & $80^{+300}_{-70}$  & $-1.5^{+0.2}_{-0.2}$  & $130^{+5}_{-20}$  \\ 
   $1000<\rm FWHM<3000$  &     $<0.21$  &     $>4.3$  &     $>2.1$  &    $-1.4^{+0.9}_{-0.2}$  &  $8.8^{+0.2}_{-0.1}$  & ---  & $-1.4^{+0.2}_{-0.2}$  & $120^{+30}_{-20}$  \\ 
 \hline
\end{tabular}
\label{stack_results}
\end{table*}


We perform the same analysis for sources in which we detect at least two other UV emission lines in addition to Ly$\alpha$, allowing for line ratios between those three or more lines (and constraints on the others) to be used. It is worth noting that requiring at least three UV lines to be detected is expected to select sources with the highest ionisation parameters, AGN nature and/or high effective temperatures. 

In general, we find that our AGN-classified sources have line-ratios which are only reproducible with power-law models or black-bodies with high temperatures around $\sim120-140$\,kK, particularly due to their N{\sc v} and He{\sc ii} fluxes. These sources have line ratios which imply ionisation parameters ($\log\,U$) from $\approx-2$ to $\approx1$ and gas-phase metallicities between $\approx0.4$\,$Z_{\odot}$ and $\approx2$\,$Z_{\odot}$, mostly driven by the strong N{\sc v} detections \citep[see e.g.][]{Matsuoka2009}, but also by C{\sc iv}. These results show that there is a diversity of properties within the AGN population of bright LAEs at least in terms of gas metallicities and ionisation parameters. The derived properties of the AGN in our sample are also very similar to those derived for high redshift radio galaxies \citep[see][]{Miley2008A&AR}.

For the only source with at least two UV lines which is likely dominated by star-formation (GN-NB5-6712), we find that BPASS models are able to reproduce the line ratios, although black body models with temperatures $\approx70-100$\,kK also reproduce the constraints on the line ratios. The {\sc cloudy}-BPASS models suggest a $\sim20$\,Myr-old burst, an ionisation parameter of $\log\,U=-3.0^{+1.7}_{-0.8}$ and sub-solar to solar gas-phase metallicity ($0.6^{+0.8}_{-0.4}$\,$Z_{\odot}$), and thus similar to properties derived for faint LAEs \citep[e.g.][]{NakajimaOuchi2013,Suzuki2017}. We note, nonetheless, that GN-NB5-6712 is not necessarily representative of the population of bright LAEs dominated by star-formation, particularly due to the detection of both C{\sc iv} and He{\sc ii}.

\subsection{Results from stacks}\label{STACK_results}

For a fraction of the LAEs the physical parameters are unconstrained due to many lines being below the detection limit. This can in principle be addressed through stacking similar sources, allowing us to better constrain the general nature of the sources within the population of luminous LAEs as a whole. We stack the sub-samples listed in Table \ref{stack_results}, and constrain the emission lines following the same methodology as for individual measurements. We stack using the Ly$\alpha$ redshift. As we are only interested in obtaining the stacked flux or luminosity of each line, we bin each spectrum so that we eliminate most of the scatter in velocity offsets between Ly$\alpha$ and other lines (to avoid blurring them out) and also make doublets unresolved. In practice, we bin each spectrum to a resolution of 750\,km\,s$^{-1}$ at rest-frame 1500\,{\AA}, motivated by the largest velocity offsets in our sample. This also allows us to measure emission-line fluxes with single Gaussians for all stacks.

In order to stack the spectra, we transform flux density into luminosity density taking into account the redshift of each source and average combine spectra weighting by the inverse of the variance, as well as masking strong OH sky lines. We also obtain median stacks which provide similar results but with lower S/N.

We show our results for a range of our stacking analysis as a function of Ly$\alpha$ luminosity in Figure \ref{1D_stack_all} and the full results in Table \ref{stack_results}. We find that the strength of rest-frame UV lines roughly scales with Ly$\alpha$ and UV luminosities (mostly driven by AGN activity), but that the dependence with increasing Ly$\alpha$ FWHM appears different, particularly due to the spectra of the broadest line emitters which show relatively symmetric Ly$\alpha$ (implying an average line-of-sight volume with very low column density), symmetric strong Ly$\beta$ and broad high ionisation metal lines like N{\sc v}, C{\sc iv} and C{\sc iii}] but weak to undetectable e.g. He{\sc ii}. The most luminous LAEs are also those with the most symmetric Ly$\alpha$ lines and show the strongest C{\sc iv}, He{\sc ii} and C{\sc iii}] lines, along with significantly blue-shifted ISM lines likely tracing strong outflows of $\sim750-1000$\,km\,s$^{-1}$ \citep[see also][]{Erb2015}; see Figure \ref{1D_stack_all}.

We find that the ionisation parameter increases with increasing Ly$\alpha$ and UV luminosities, being the highest for the stack of AGN sources ($\log\,U=0.6\pm0.5$) and the lowest for SFGs ($\log\,U=-3.0^{+1.6}_{-0.9}$). It is therefore likely that trends with luminosity are driven by the prevalence of AGN sources at bright luminosities and at higher FWHMs (see Figures \ref{AGN_FRACTIONS} and \ref{MUV_andLLya_vs_LyaFWHM}) and not due to a change in properties of star-forming dominated LAEs. We also find evidence for gas-phase metallicities of LAEs to be lower ($\approx0.1-0.3$\,Z$_{\odot}$) for SFGs (see Table \ref{stack_results}) and for lower luminosity LAEs \citep[similar to e.g.][]{Stark2014,Nakajima2018}, and closer to solar or higher for the most luminous LAEs (Table \ref{stack_results}), but again this is likely caused by the sharp transition between star-forming and AGN-dominated (see Figure \ref{AGN_FRACTIONS}).

\section{Discussion} \label{discussion}

\subsection{Bright LAEs at $\bf z\sim2-3$: the SF-AGN transition and the physical interpretation}\label{disc_AGN_SF}

Our results show that the AGN fraction of luminous LAEs strongly depends on both UV and Ly$\alpha$ luminosity at $z\sim2-3$. The brightest LAEs in Ly$\alpha$ or in the UV are AGN, and above $\rm L_{Ly\alpha}>10^{43.3}$\,erg\,s$^{-1}$ or $\rm M_{UV}<-21.5$ virtually all luminous LAEs are AGN, causing an abrupt change in the physical properties of the ionised gas at $2\times$\,L$^*$, as seen in \S\ref{STACK_results}. This may be explained by the fact that only AGN can reach the highest observed luminosities in either UV or Ly$\alpha$, and suggests a limiting observed SFR\footnote{Obtained converting the observed maximal $\rm M_{UV}=-21.5$ or $\rm L_{Ly\alpha}=10^{43.3}$\,erg\,s$^{-1}$ to a SFR using a Salpeter IMF.} for star-forming dominated LAEs of $\approx20$\,M$_{\odot}$\,yr$^{-1}$ at $z\sim2-3$. In principle, such a SFR limit could be a consequence of the exponential cut-offs in the galaxy mass and the gas mass fraction functions, and the properties of dust formation in massive starbursts, with both effects potentially evolving with redshift (see \S\ref{AGN_SF_evolution}). These likely combine to create a strong threshold on UV luminosity of SF-dominated systems.

If dust is the main driver of a sharp UV and Ly$\alpha$ observed luminosity limit for starbursts (leading to a sharp transition between SF and AGN dominated at $2\times$\,L$^*$), one would expect the limit to be different when looking at H$\alpha$ luminosity of H$\alpha$ selected sources, which is significantly less affected by dust. Results from \cite{Sobral2016} show that there is a rise in the AGN fraction of H$\alpha$ selected sources as a function of H$\alpha$ luminosity above $2\times$\,L$^*_{\rm H\alpha}$, but such rise is much slower and not as abrupt as the one found for LAEs (see Figure \ref{AGN_FRACTIONS}) with Ly$\alpha$ or UV luminosities. H$\alpha$ emitters only become 100\% AGN dominated at an observed H$\alpha$ luminosity of about 10$^{43.8}$\,erg\,s$^{-1}$, corresponding to an observed SFR limit of $\approx500$\,M$_{\odot}$\,yr$^{-1}$ in H$\alpha$ luminosity, 25 times larger than the limit for the observed UV and Ly$\alpha$ SFRs for LAEs. We therefore interpret the sharp shift from star-forming to AGN dominated sources at $2\times$\,L$^*$ for LAEs as likely a combination of two effects: a rise in the AGN fraction with intrinsic luminosity \citep{Sobral2016}, as seen for H$\alpha$ selected samples, and the dominating factor of increased obscured SFRs as a function of increasing intrinsic SFR \citep[e.g.][]{GarnBest2010,Swinbank2004,Whitaker2017}, leading to a maximal unobscured starburst. 

While intrinsically star-formation dominated sources can still reach even higher SFRs and produce higher UV and Ly$\alpha$ luminosities, such intense starbursts happen in dusty environments or starbursts themselves produce the dust at $z\sim2-3$ \citep[showing up as dusty star-forming galaxies, see][]{Casey2014_REV}, thus limiting or even reducing the observed UV and Ly$\alpha$ luminosities. The high intrinsic production of UV and Ly$\alpha$ photons of dusty star-forming galaxies suggests that any escape channel created by intense feedback in those galaxies will still lead to non-negligible UV and Ly$\alpha$ luminosities which are observed in a variety of studies \citep[see e.g.][]{Chapman2005,NILSSON2009LL,Casey2014,Matthee2016}, but that given our results will not exceed $2\times$\,L$^*$ at $z\sim2-3$.

We conclude that the clear separation between star-forming dominated and AGN dominated LAEs at $2\times$\,L$^*$ is caused primarily by obscured star-formation becoming dominant above observed SFRs of $\approx20$\,M$_{\odot}$\,yr$^{-1}$. Prominent dust obscuration will efficiently absorb UV photons and directly or indirectly (via scattering) lead to low Ly$\alpha$ escape fractions \citep[e.g.][]{Atek2008,Matthee2016} such that the observed luminosities saturate. Ly$\alpha$ and UV luminosities observed above $2\times$\,L$^*$ are thus likely to be produced by different physical processes linked with AGN activity, which in principle do not create dust. These may include strong AGN-driven outflows and radiation from different regions within the accretion disk, as seen from the evidence of the UV bright broad line emitters, while shocks may also play a role. The existence of UV blue, Ly$\alpha$ bright quasars \citep[e.g.][]{Borisova2016} implies that the physical processes in accretion disks of massive black holes are scalable to much higher UV luminosities without dust imposing a strong limitation in the observed luminosities. This is likely due to outflows and high ionisation radiation fields which we find to be be present ($\log\,U$ from $\approx-2$ to $\approx1$; see \S\ref{CLOUDY_lines_AGN_SF} and \S\ref{STACK_results}) that may destroy dust or open channels for Ly$\alpha$ and UV photons to escape \citep[see also][]{Venemans2007,Miley2008A&AR}. Nevertheless, it should be noted that our most luminous LAEs already have Ly$\alpha$ luminosities as high \citep[e.g.][]{Borisova2016} as the ones found for the most UV luminous quasars \citep[e.g.][]{Richards2006}, and reaching down to luminosities below those of fainter AGN \citep[e.g.][]{Gavignaud2006}. Therefore, while the UV luminosity seems to be scalable up from even our brightest UV sources by a few orders of magnitude, there may be a saturation in the observable Ly$\alpha$ luminosity at close to $10^{44-45}$\,erg\,s$^{-1}$ for AGN at $z\sim2-3$.

\subsection{Implications for higher redshift: evolution?}\label{AGN_SF_evolution}

Our results and physical interpretation at $z\sim2-3$ may allow us to shed further light into the nature and evolution of higher redshift LAEs selected in the same way. Recently, \cite{Sobral2017_SC4K} obtained a large sample of $\sim4000$ LAEs from $z\sim2$ to $z\sim6$ and explored it to infer a strong evolution (a factor of $\approx4-5$) in the typical Ly$\alpha$ escape fraction of star-forming galaxies, rising as $(1+z)^2$. This increase is in addition to the likely increase of $\rm \xi_{ion}$ (tracing high burstiness and/or an average change in stellar populations) by a factor of $\sim2$, leading to a total observed rise of $(1+z)^3$ in the Ly$\alpha$ to UV luminosity density ratio with increasing redshift \citep[see also][]{Hayes2011}. Furthermore, \cite{Sobral2017_SC4K} finds a factor $\sim5$ rise in L$^*_{\rm Ly\alpha}$ from $z\sim2$ to $z\sim6$, which, under our interpretation of $2\times$\,L$^*_{\rm Ly\alpha}$ being the limit of the observed Ly$\alpha$ luminosity of a starburst would imply that by $z\sim6$ the limiting observable Ly$\alpha$ luminosity for a starburst will be $\approx10^{44}$\,erg\,s$^{-1}$. If this is the case, by $z\sim6$ the limiting unobscured SFR would be $\approx100$\,M$_{\odot}$\,yr$^{-1}$, although the UV observed SFR may be limited to a lower value due to the high ratio of LyC to UV photons ($\rm \xi_{ion}$) expected by $z\sim6$ \citep[see e.g.][]{Matthee2017}. These results would provide a natural explanation for the exceptionally high Ly$\alpha$ luminosities of equally selected LAEs at $z\sim6-7$ \citep[][]{Ouchi2009,Sobral2015,Hu2016,Matthee2017c} with $\rm L_{Ly\alpha}\approx10^{43.4-43.9}$\,erg\,s$^{-1}$ that so far show no convincing evidence for being dominated by AGN activity \citep[see also][]{Pallottini2015,Mas-Ribas2016,Bagley2017,Pacucc2017}. For example, these luminous $z\sim6-7$ LAEs show Ly$\alpha$ profiles which are typically narrow ($\approx200-300$\,km\,s$^{-1}$; \citealt{Matthee2017c}), contrarily to the high FWHMs we find at $z\sim2-3$ for the same luminosities (see Figure \ref{MUV_andLLya_vs_LyaFWHM}). According to our interpretation of an evolving L$^*_{\rm Ly\alpha}$ by a factor of 5, the $z\sim6-7$ bright LAEs would easily be consistent with being star-forming dominated, with CR7's Ly$\alpha$ luminosity \citep[$\approx10^{43.9}$\,\,erg\,s$^{-1}$;][]{Sobral2015} being the closest to our inferred maximal unobscured starburst luminosity of $\approx10^{44}$\,erg\,s$^{-1}$ at $z\sim6$. Recent deep ALMA observations of CR7 \citep{Matthee2017_ALMA} are fully consistent with this interpretation, with CR7's obscured SFR being below $<4$\,M$_{\odot}$\,yr$^{-1}$.

Potential mechanisms which may be able to raise the maximal unobscured luminosity of early luminous LAEs are (major) mergers (seen as a single source) and/or significant gas inflows without reducing the Ly$\alpha$ escape fraction and/or without leading to significant dust obscuration. The necessary conditions likely require dust destruction and/or inefficient dust formation at low metallicities \citep[see e.g.][]{Behrens2018}, which may be more common at higher redshift, including the possibility of witnessing a first significant burst of star-formation. Interestingly, there is evidence for the brightest LAEs at $z\sim7$ being multiple component/mergers \citep[][]{Ouchi2010,Sobral2015,Sobral2017_CR7,Matthee2017c}, while those at $z\sim$\,2--3 are typically very compact and single component dominated \citep[e.g.][]{Paulino-Afonso2017_LAEs}. At $z\sim2-3$ major mergers lead to very high SFRs but that are significantly obscured \citep[see][]{Casey2014_REV}.

Our scenario of an evolving maximal unobscured star-burst is consistent with results from \cite{Calhau2018} who find that the relation between the X-ray AGN fraction of LAEs and Ly$\alpha$ luminosity declines with redshift or shifts to higher luminosities at higher redshift. Our discussed scenario, along with its physical implications, can easily be tested with the {\it James Webb Space Telescope (JWST)}. Using {\it JWST}'s IFU capabilities in the NIR and MIR regimes one can observe CR7, Himiko, VR7, MASOSA and other luminous LAEs \citep[see e.g.][]{Matthee2017c} and establish them as either star-forming or AGN dominated, along with measuring their resolved dust obscuration using the Balmer decrement. Most interestingly, IFU data will allow to start understanding the physical reasons behind a potentially higher maximal unobscured starburst at high redshift and implications for dust formation and stellar populations in early luminous LAEs.

\section{Conclusions} \label{conclusions}

We have presented the spectroscopic follow-up of $23$ luminous ($\rm L_{Ly\alpha}\gsim L^*_{Ly\alpha}$) Ly$\alpha$ candidates at $z\sim$\,2--3 found with large narrow-band surveys in Bo\"{o}tes, COSMOS, GOODS-N, SA22 and UDS. We have used WHT/ISIS, Keck/DEIMOS and VLT/X-SHOOTER to spectroscopically confirm 21 of them as luminous LAEs. We exploit the wide wavelength coverage of our data and the high resolution spectra to measure line profiles, investigate the nature of the sources and explore UV emission line ratios. Our main results are:

\begin{itemize}

\item Luminous LAEs at $z\sim$\,2--3 present a striking diversity in terms of their Ly$\alpha$ and UV properties. They also present a wide diversity in the shape of the Ly$\alpha$ line, ranging from roughly symmetric to highly asymmetric and including 25\% of sources with significant blue-shifted Ly$\alpha$ components or double-peaked lines, typically linked with high Ly$\alpha$ escape fractions. 

\item Luminous LAEs are in general very blue (UV slope of $\beta=-2.0^{+0.3}_{-0.1}$) and span five orders of magnitude in rest-frame UV luminosities, from $\sim0.3$\,M$^*_{\rm UV}$ to $\sim30$\,M$^*_{\rm UV}$. They likely have a low dust content, consistent with $E(B-V)\approx0.05$ and have high Ly$\alpha$ escape fractions of $50^{+20}_{-15}$\,\% after correcting for dust extinction.

\item Besides Ly$\alpha$, the most prevalent high ionisation UV line in luminous LAEs is C{\sc iv}, present in 40\% of the sample, but we also find N{\sc v}, He{\sc ii}, C{\sc iii}] and O{\sc iii}] as relatively common lines, particularly at the brightest UV and Ly$\alpha$ luminosities.

\item Ly$\alpha$ FWHMs of luminous LAEs vary from $\sim200$ to $\sim4000$\,km\,s$^{-1}$, with the prevalence of broad lines rising significantly with increasing Ly$\alpha$ and UV luminosities. We find significant relations between the Ly$\alpha$ FWHM and Ly$\alpha$ and UV luminosities for NL-AGN, likely driven by outflows.

\item The narrow LAEs have small velocity offsets between Ly$\alpha$ and other lines of only $\sim100-200$\,km\,s$^{-1}$, consistent with Ly$\alpha$ photons escaping relatively close to systemic and providing further evidence for high Ly$\alpha$ escape fractions in bright LAEs.

\item We find a significant increase in the AGN fraction of LAEs with both Ly$\alpha$ and UV luminosities from $\sim0$ to $\sim100$\%. Above $\rm L_{\rm Ly\alpha}>10^{43.3}$\,erg\,s$^{-1}$ and/or $\rm M_{UV}<-21.5$ all luminous LAEs at $z\sim2-3$ are AGN. We find simple relations to predict the AGN fraction of LAEs (within the ranges studied): $\rm f_{AGN}=(-0.30\pm0.07)({\rm M_{UV}+20.5})+(0.35\pm0.11)$ and $\rm f_{AGN}=(0.78\pm0.22){(\rm \log_{10}(L_{Ly\alpha})-42.72)}+(0.24\pm0.14)$.

\item The transition from star-forming dominated to AGN dominated happens at $2\times$\,L$^*$ and is likely linked with a physical limit for the maximum observed output of an unobscured starburst of $\approx20$\,M$_{\odot}$\,yr$^{-1}$ at $z\sim2-3$. For higher intrinsic UV and Ly$\alpha$ luminosities dust likely prevents such sources from being observed as brighter than $2\times$\,L$^*$.

\item AGN LAEs reveal high ionisation parameters ($\log U=0.6\pm0.5$) and sub-solar to solar metallicities. Luminous LAEs lacking signatures of AGN ($40\pm11$\,\%) are less ionising ($\log U=-3.0^{+1.6}_{-0.9}$ and $\log\xi_{\rm ion}=25.4\pm0.2$), apparently metal-poor (12+log(O/H)$=8.2^{+0.5}_{-0.3}$) sources likely powered by young, dust-poor ``maximal" starbursts.

\item The ionisation parameter rises with Ly$\alpha$ and UV luminosities by typically 1\,dex from the low to high luminosities, while gas-phase metallicities rise with Ly$\alpha$ luminosity from 12+log(O/H)$=7.5^{+0.5}_{-0.1}$ to 12+log(O/H)$=8.7^{+0.1}_{-0.1}$. We interpret these results as a consequence of the transition between star-forming dominated systems to AGN dominated which we find to be particularly abrupt at $2\times$\,L$^*$.

\end{itemize}

\section*{Acknowledgments}

We thank the anonymous reviewer for their timely and constructive comments that greatly helped us to improve the manuscript. DS acknowledges financial support from the Netherlands Organisation for Scientific research (NWO) through a Veni fellowship and from Lancaster University through an Early Career Internal Grant A100679. JM acknowledges the support of a Huygens PhD fellowship from Leiden University. BD acknowledges financial support from NASA through the Astrophysics Data Analysis Program (ADAP), grant number NNX12AE20G, and the National Science Foundation, grant number 1716907. IRS acknowledges support from the ERC Advanced Grant DUSTYGAL (321334), STFC (ST/P000541/1) and a Royal Society/Wolfson Merit Award. PNB is grateful for support from STFC via grant ST/M001229/1. We thank Anne Verhamme, Kimihiko Nakajima, Ryan Trainor, Sangeeta Malhotra, Max Gronke, James Rhoads, Fang Xia An, Matthew Hayes, Takashi Kojima, Mark Dijkstra, and Anne Jaskot for many helpful and engaging discussions, particularly during the SnowCLAW Ly$\alpha$ workshop. We thank Bruno Ribeiro, Stephane Charlot and Joseph Caruana for comments on the manuscript. The authors would also like to thank Ingrid Tengs, Meg Singleton, Ali Khostovan and Sara Perez for participating in part of the observations. We also thank Jo\~ao Calhau, Leah Morabito, S\'ergio Santos and Aayush Saxena for their assistance with the narrow-band observations which allowed to select some of the sources. Based on observations obtained with the William Herschel Telescope (WHT), program: W16AN004 and with Keck, program C267D. Based on data products from observations made with ESO Telescopes at the La Silla Paranal Observatory under ESO programme IDs 294.A-5018, 294.A-5039, 092.A-0786, 093.A-0561, 097.A-0943, 098.A-0819, 099.A-0254 and 179.A-2005. The authors acknowledge the award of service time (SW2014b20) on the WHT. WHT and its service programme are operated on the island of La Palma by the Isaac Newton Group in the Spanish Observatorio del Roque de los Muchachos of the Instituto de Astrofisica de Canarias. The authors would also like to thank all the extremely helpful observatory staff that have greatly contributed towards our observations, particularly Fiona Riddick, Lilian Dominguez, Florencia Jimenez and Ian Skillen.

We have benefited greatly from the publicly available programming language {\sc Python}, including the {\sc NumPy} \& {\sc SciPy} \citep[][]{van2011numpy,jones}, {\sc Matplotlib} \citep[][]{Hunter:2007}, {\sc Astropy} \citep[][]{Astropy2013} and the {\sc Topcat} analysis program \citep{Topcat}. This research has made use of the VizieR catalogue access tool, CDS, Strasbourg, France.

The samples and parent samples used for this paper are publicly available \citep[][]{Sobral2017,Sobral2017_SC4K,Matthee2017b}. We also make all our 1D (binned to one third of the resolution, flux calibrated, without any slit correction) spectra public on-line with the final refereed version of this paper, together with the catalogue of luminous LAEs and electronic versions of relevant tables (see Appendix \ref{reduced_2D_1D}).

\bibliographystyle{mnras}
\bibliography{bib_LAEevo.bib}

\appendix

\section{Observations and Data reduction}\label{WHIS_pipeline}

Tables \ref{Obs_table_ISIS}, \ref{Obs_table_DEIMOS} and \ref{Obs_table_XSHOOT} present the observing logs for WHT/ISIS, Keck/DEIMOS and VLT/X-SHOOTER observations, respectively. In Table \ref{noise_resolution_final} we present the best estimates of the spectral and noise properties of all reduced data for WHT/ISIS, Keck/DEIMOS and VLT/X-SHOOTER.

%
%
\begin{table*}
\caption{Observing log for WHT/ISIS (as part of W16AN004, PI: Sobral) of our luminous LAEs at $z\sim2$ \citep[from][]{Sobral2017} and $z\sim$\,2--3 \citep[from][]{Matthee2017b} observed in the Bo\"{o}tes and GOODSN fields. We order sources based on the predicted Ly$\alpha$ luminosity from the corresponding narrow-band estimate (brightest on top, faintest at the bottom). Exposure times (ET) are the same for the blue and red arms, and thus we do not give them individually. *BR3 and BH-NB5-16 were also observed with Keck/DEIMOS. All the sources were confirmed as real LAEs at either $z\sim3.1$ or $z\sim2.2$, although note that BH-NB392-55 does not have Ly$\alpha$ coverage in the observational set-up and it was confirmed by the strong presence of C{\sc iv}; see e.g. Figure \ref{2D_full_ISIS}.}
\label{Obs_table_ISIS}
\begin{tabular}{ccccccccc}
\hline
Target	& Grism  &	 ET (target)	& ET (star)  & Calib Star	 & Date  & Seeing & Sky & Moon  \\
 (ID)   &  &   (ks) 	&	(ks)	 & (SDSS ID)  & (2016)  &($''$)	& & \\
\hline 	
BH-NB5-4 (BR3)* & R600 & 3.0  & 0.1 & J143232.23+333903.7 &    5 May & 0.9  & Clear & Dark  \\   
BH-NB5-6 & R300/R316 &5.4  & 0.2 &  J143317.66+335348.9  &    1 Jul & 0.8  & Clear & Dark  \\  
BH-NB392-12 & R600 & 3.0 & 0.1 &	J143026.70+333455.4 &  4 May  &	1.0  & Clear & Dark  \\   
BH-NB5-16* & R300/R316 & 6.3  & 0.2 &	J143102.13+340329.1  &    30 Jun	&	0.6  & Clear & Dark  \\  
GN-NB501-6712 & R600 & 11.5  & 0.4  &	J123617.51+622445.9  &   4-5 May   &	1.0  & Clear & Dark  \\   
BH-NB392-55  & R600 & 4.0 & 0.1 & J143043.56+340349.4 &    5 May	 &	1.0  & Clear & Dark  \\   
\hline 
\end{tabular}
\end{table*}

\begin{figure*}
\includegraphics[width=14.6cm]{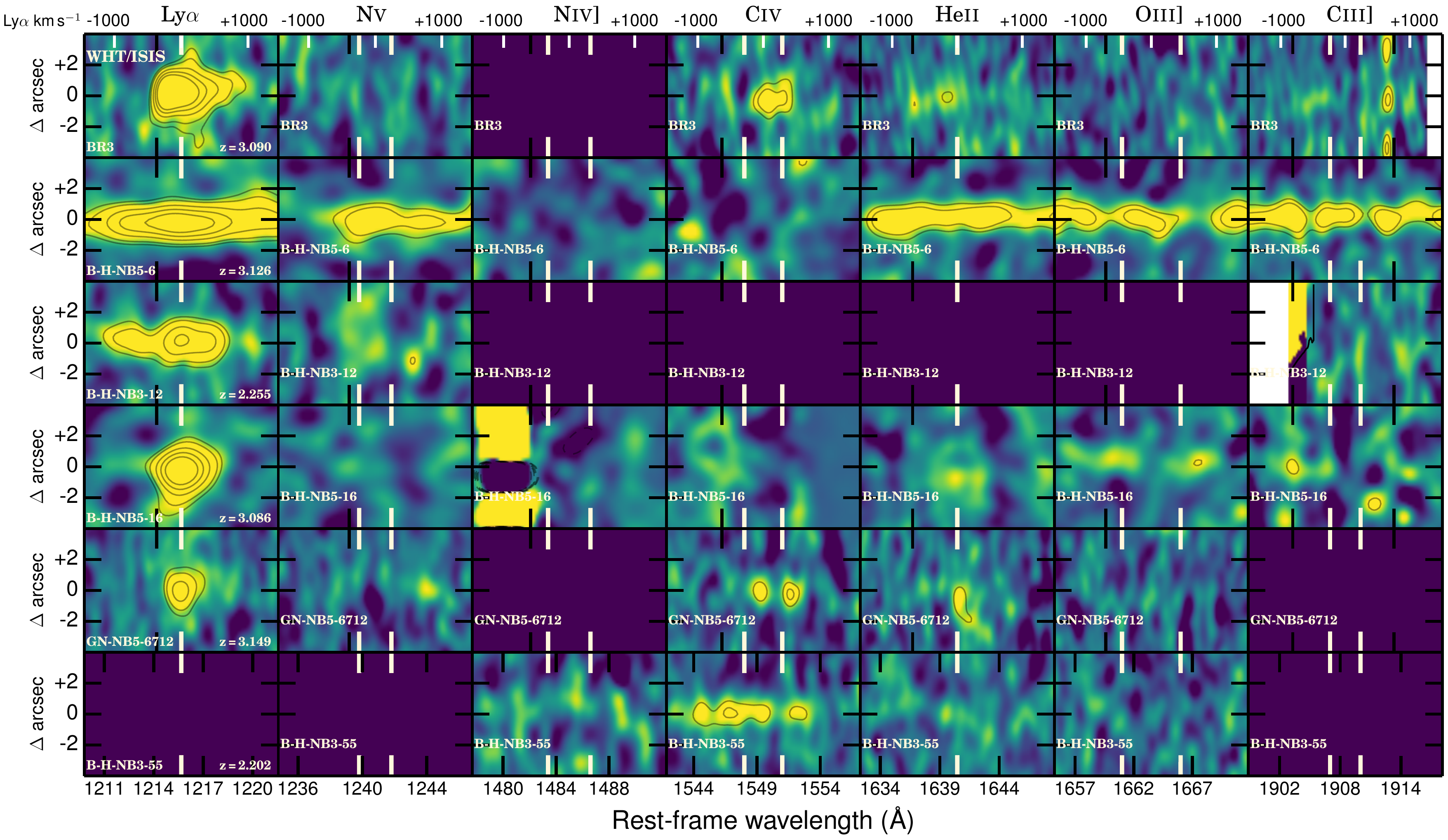}
\caption{Reduced 2D spectra from our WHT/ISIS observations and zoom in into the major emission lines studied here. We order sources according to their Ly$\alpha$ luminosity, from the more luminous at the top, to the faintest at the bottom. We use low and high cut-offs corresponding to $-1$\,$\sigma$ and $+2$\,$\sigma$. We show contours corresponding to $1.5,2,3,4,5,10$\,$\sigma$ in all 2D zoom-ins. Velocity offsets are given with respect to Ly$\alpha$. Windows of the 2D spectra in dark represent spectral regions without spectral coverage, where no information about the specific line is available for the specific instrument/source combination. Apart from significant Ly$\alpha$ detections, we find one source with strong N{\sc v} (BH-NB5-6), three sources with C{\sc iv}, another three sources with He{\sc ii} and 1-2 sources with O{\sc iii}] or C{\sc iii}]. GN-NB5-6712 is particularly interesting, showing redshifted C{\sc iv} emission (relative to Ly$\alpha$) and no He{\sc ii} offset; this could mean Ly$\alpha$ is escaping roughly at systemic, with C{\sc iv} likely indicating that this is due to an outflow likely clearing the path to Ly$\alpha$ photons and potentially LyC photons (Sobral et al. in prep.).}
\label{2D_full_ISIS}
\end{figure*}

\subsection{A {\sc python} WHT/ISIS pipeline: WHIS}\label{WHIS_pipeline}

Here we present and explain in detail our {\sc python} WHT/ISIS data reduction pipeline. On a night by night basis, the pipeline starts by identifying the different calibration and science frames, and groups them appropriately. Bias frames are median combined to produce a master bias. All frames are then bias subtracted. Flats within the linear regime are combined, per arm, grating and central wavelength, after normalising each flat by its median, and then median combined to produce different normalised master flats. All individual raw frames are then flat fielded using the appropriate normalised master flat. Arcs are median combined per arm, grating and central wavelength to produce master arcs. We combine arcs separately for the morning and evening arcs because we notice that small shifts can happen across the night.

%
 %
\begin{table}
 \centering
\caption{Final spectral and median noise properties of the binned (typically binned by $\sim2$ pixels) spectra ($\pm100$\,{\AA} of $\lambda$ given and excluding strong OH lines) of all reduced data for WHT/ISIS, Keck/DEIMOS and VLT/X-SHOOTER. The 1$\sigma$ noise levels are in units of erg\,s$^{-1}$\,cm$^{-2}$\,{\AA}$^{-1}$. Note that in the case of X-SHOOTER, the noise in the NIR arm can vary significantly at the position of the strongest sky lines.} 
 \label{noise_resolution_final}
\begin{tabular}{ccccc}
\hline
Telescope/   & Arm/  & $\lambda$ & FWHM & 1\,$\sigma$ noise \\ 
 Instrument  &  Grism  &  ({nm}) & (km\,s$^{-1}$) & ($\times10^{-18}$) \\ 
\hline
WHT/ISIS & R600B & 400 & 150 & 15 \\
WHT/ISIS & R600R & 500 & 120  & 8 \\
WHT/ISIS & R600R & 700 & 77 & 4 \\
WHT/ISIS & R300B & 500 & 240 & 2 \\
WHT/ISIS & R316R & 700 & 154 & 1 \\
\hline
Keck/DEIMOS & B 600L & 500 & 180 & 0.5 \\
Keck/DEIMOS & R 600L & 700 & 130 & 0.3 \\
\hline
VLT/XSHOOT & UVB & 400 & 120 & 6 \\
VLT/XSHOOT & UVB & 500 & 96 & 3 \\
VLT/XSHOOT & VIS & 700 & 69 &  5 \\
VLT/XSHOOT & NIR & 1600 & 67 & 3 \\
\hline
\end{tabular}
\end{table}

Master arcs are used to extract the 1D arc spectra per night, per grating and central wavelength. Arcs lines are matched to air wavelength CU+CAr lines, matching $\sim30$ spectral lines covering the full spectral range per arm. We fit 3rd order polynomials and find the rms in the wavelength calibration solution to always be between 0.1-0.2\,\AA \ without any noticeable trend. We perform a final check and, if needed, a small correction to the wavelength calibration on a source by source basis (typically $<0.5$\,\AA) by exploring the available sky lines obtained from the same frames as the science targets, allowing to trace any small changes.

For each science target, we start by using the corresponding acquisition star frames and bin the 2D spectrally and spatially in order to automatically find the trace of the star in each individual 2D spectra. We find that fitting a line to the trace is able to trace it without any offsets along the wavelength direction. We then use the position of the trace in the spectral centre of the 2D as the centre of the trace for the target. We use those to find the exact location of the trace in the multiply offset spectra and to compute the required offsets in pixels. We use these to subtract both star and science spectra in sets of ABBA and to combine them. We then obtain an average stack (we also obtain median stacks and stacks without sky subtraction), removing individual pixels which are more then 5$\sigma$ away from the average counts, to obtain a final average stacked 2D of the science target and its appropriate star. We then automatically find the trace in the star 2D spectrum and use it to extract both the science and the star spectra. We also extract the sky spectrum. We use the sky spectra to check the wavelength calibration and to apply any small wavelength offsets needed based on the calibrated ISIS spectra of OH lines. As a final sky subtraction step, we minimise residuals from sky lines by measuring the average counts away from the location of the source (where the average should be zero), and subtract those counts per resolution element. We show 2D spectra in Figure \ref{2D_full_ISIS}.

We flux calibrate 1D spectra by using the calibration star and taking full advantage of our observation methodology (of always observing a calibration star prior to each science exposure). We use SDSS magnitudes to produce an artificial spectrum of each star. We then compute the ratio between total counts measured and the flux densities derived from SDSS for each calibration star. We do this after masking all positions of strong lines (e.g. Ca HK, H$\delta$, H$\beta$, H$\alpha$) and linearly interpolate between those wavelengths. We find the flux calibration to be accurate within $\sim10$\%. By definition, due to the way we extract our science targets (using the trace of the stars which we use for flux calibration) all point sources will be slit corrected. We use our narrow-band imaging to comment on the need for any extra corrections for some of our sources which are not point-like. Our spectra have a typical rms/depth of $1-2\times10^{-17}$\,erg\,s$^{-1}$ for a 200\,km\,s$^{-1}$ line (rest-frame) in both blue and red arms. We show examples of 1D extracted spectra in Figure \ref{1D_examples_fitting}.

%
%
\begin{table*}
\caption{Observing log for the different 3 observing nights with Keck DEIMOS (as part of program C267D; PI Darvish) for luminous $z\sim3$ LAEs in the Bo\"{o}tes and GOODS-N fields from \citet{Matthee2017b}. Observing nights of 3 June 2016 and 7 July 2016 used a dither pattern of $\pm2.5''$, while for the 30 July 2016 run no dithering was applied. The 3 sources after the main 8 targets were targeted as part of C267D out of the sample of all emission line candidates (ELC) which were not Ly$\alpha$ selected; one turned out to be a real LAE, while the other two are powerful quasars, selected due to significant transmission at $\sim5000$\,{\AA}. We also provide information and the reduced spectra for these final 3 sources, but due to the different selection we do not include them in the main analysis of Ly$\alpha$ selected sources. The final 2 sources were targeted initially as potential bright Ly$\alpha$ candidates, and then confirmed to be [O{\sc ii}] emitters at $z\sim0.3$, but were not included in \citet{Matthee2017b} as LAE candidates; we list them here for completeness. Feige 66 was used as a flux calibrator along with a source with bright enough continuum, B-NB921-198 \citep{Matthee2017b}. **For BH-NB5-34 the Ly$\alpha$ coverage is partially missing along a spectral-spatial direction.}
\label{Obs_table_DEIMOS}
\begin{tabular}{cccccccc}
\hline
Target  & Mask 	&	 ET (targets)& Date  & Seeing & Sky & Moon &   Nature \\
(ID)  & (name)   &  (ks)   	&	 (2016)  & ($''$)	& & & ($z_{spec}$ and emission line)  \\
\hline
BH-NB5-4 (BR3)*  & btm-1 &   5.4  & 3 June  &  0.7  & Clear & Dark & $z=3.1$ LAE   \\  
BH-NB5-10  & btm-1 &   5.4  &  3 Jun  &  0.7  & Clear & Dark  & $z=3.1$ LAE     \\  
BH-NB5-16*  & btm-5 &   7.2  &   30 Jul	&	0.6 &  Clear & Dark  &  $z=3.1$ LAE    \\      
BH-NB5-27 & btm-2 &  4.5 &  3 Jun  &	 0.6  & Clear & Dark  &  $z=3.1$ LAE 	\\ 
BH-NB5-34** & btm-3 &    5.4 &  7 Jul  &	0.8 & Clear & Dark &  $z=3.1$ LAE  	\\  
BH-NB5-37  & btm-1 &   5.4  &  3 Jun  &  0.7  & Clear & Dark  & $z=3.1$ LAE      \\  
GN-NB501-3378 & gdsn & 7.2  &    3 Jun	 &	1.0  & Clear & Dark & $z=3.2$ LAE     \\  
GN-NB501-5878 & gdsn & 7.2  &   3 Jun	 &	1.0  & Clear & Dark & $z=3.1$ LAE    \\   
\hline
B-NB501-31679 & btm-1 &  4.5 &   3 Jun  &	0.6  & Clear & Dark & ELC  $z\sim3$ quasar  (Ly$\alpha$)	\\ 
B-NB501-25550 & btm-1 &  4.5 &   3 Jun  &	0.6  & Clear & Dark & ELC  $z\sim4$ quasar  (Ly$\alpha$ forest selected)  	\\ 
B-NB501-24390 & btm-1 &  4.5 &   3 Jun  &	0.6  & Clear & Dark   & ELC  $z=3.1$ Ly$\alpha$	\\ 
\hline
BH-NB5-9 & btm-1 &  4.5 &   3 Jun  &	0.6  & Clear & Dark   & $z=0.3$ [O{\sc ii}] emitter	\\ 
BH-NB5-13  & btm-1 &  4.5 &   7 Jul  &	0.6  & Clear & Dark   & $z=0.3$ [O{\sc ii}] emitter	\\ 
\hline 
\end{tabular}
\end{table*}

\begin{figure*}
\includegraphics[width=17.6cm]{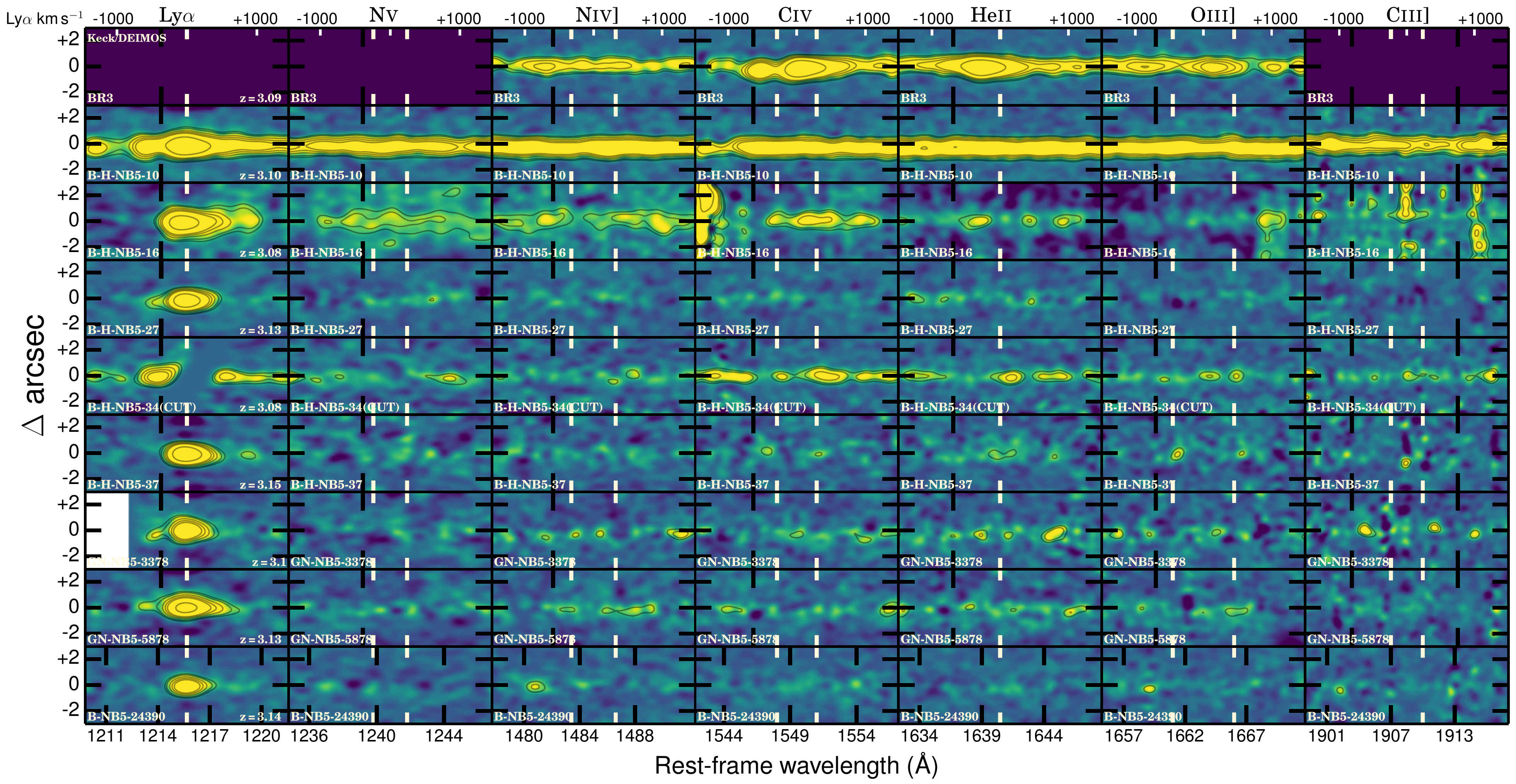}
\caption{Reduced 2D spectra from our DEIMOS observations and zoom in into the major emission lines studied here. Sources are ordered by decreasing Ly$\alpha$ luminosity (top to bottom). We use low and high cut-offs corresponding to $-1$\,$\sigma$ and $+2$\,$\sigma$. We show contours corresponding to $1.5,2,3,4,5,10,20$\,$\sigma$ in all 2D zoom-ins. Velocity offsets are given in respect to Ly$\alpha$.}
\label{2D_full_DEIMOS}
\end{figure*}

%
%
\begin{table*}
\caption{Observing log for the VLT/X-SHOOTER programs (098.A-0819 and 099.A-0254, PIs: Sobral, Matthee) for luminous LAEs at $z\sim2.2-3.1$ from \citet{Sobral2017} and \citet{Matthee2017b} in COSMOS, UDS and SA22. Sources are ordered based on the predicted Ly$\alpha$ luminosity measured from the narrow-band data, if all were LAEs. Exposure times (ET) are given separately for each X-SHOOTER arm/spectrograph as they vary slightly due to the different read-out times. The final column presents what each source was found to be and the approximate redshift using Ly$\alpha$. Overall, out of the 11 targets, we confirm nine as LAEs, and two as contaminants, with one being a strong [O{\sc ii}] emitter with no detectable continuum, and another one being a star with narrow and broad band features which mimic those of some quasars.}
\label{Obs_table_XSHOOT}
\begin{tabular}{cccccccccc}
\hline
Target 	&	 ET UVB & NIR & VIS & Calib. Star  & Dates  & Seeing & Sky & Moon & Nature \\
 (ID)   &  (ks)   	&	(ks) 	& 	(ks)  & (name)	 &   & ($''$)	& & & ($z_{spec}$) \\
\hline 	
SA22-NB5-14 & 2.4 & 2.6 & 1.6	&	Hip108612 & 28 Jun 2017 & 1.3 &  Clear& Grey &  $z=3.1$ Ly$\alpha$\\	
SA22-NB5-18 & 2.4 & 2.6  & 1.6 &	  Hip108612& 28 Jun 2017 &  1.2 &  Clear & Grey &  $z=3.1$ Ly$\alpha$ \\	
SA22-NB5-10 & 2.4 & 2.6  & 1.6 &	   F-110	&	19 Jun 2017 & 1.0  &  Clear & Grey & $z=3.1$ Ly$\alpha$  \\
CALYMHA-85 & 1.2 & 1.2  & 0.8 & Hip033300 &	  20 Jan 2017 	&	1.3 &  Thin & Grey & $z=0.0$ star \\  	
CALYMHA-383 & 0.6 & 0.6  & 0.4 & Hip033300 &    25 Oct 2016 	&	1.1 & Clear & Dark & $z=3.2$ Ly$\alpha$  \\ 	
CALYMHA-415 & 1.2 & 1.2  & 0.8 & GD71  &   4 Oct 2016	 &	1.3  & Clear & Dark & $z=2.2$ Ly$\alpha$	\\		  
CALYMHA-438 & 1.2 & 1.2  & 0.8 & Hip033300 &     25 Oct 2016	 &	1.4 & Clear & Dark & $z=0.1$ [O{\sc ii}]	\\ 
CALYMHA-373 & 12 & 12  & 8 & F-110/EG274  &   2/22/30-31 Jul 2017    &	0.9  & Clear & Dark & $z=2.2$ Ly$\alpha$  \\  
CALYMHA-67 & 1.2 & 1.2  & 0.8  & Hip033300 &	   20 Jan 2017	&	0.9 &  Thin & Grey & $z=2.2$ Ly$\alpha$  \\  
CALYMHA-147 & 12 & 12  & 8 & Hip039540 &	   20 Jan, 21-23 Dec 2017	&	0.7 &  Clear & Dark & $z=2.2$ Ly$\alpha$  \\  
CALYMHA-95 & 12 & 12  & 8 & Feige110 &	  28 Dec 2017 12-13 Jan 2018  &	0.8 &  Clear & Dark & $z=2.2$ Ly$\alpha$  \\  
\hline 
\end{tabular}
\end{table*}

\begin{figure*}
\includegraphics[width=17.6cm]{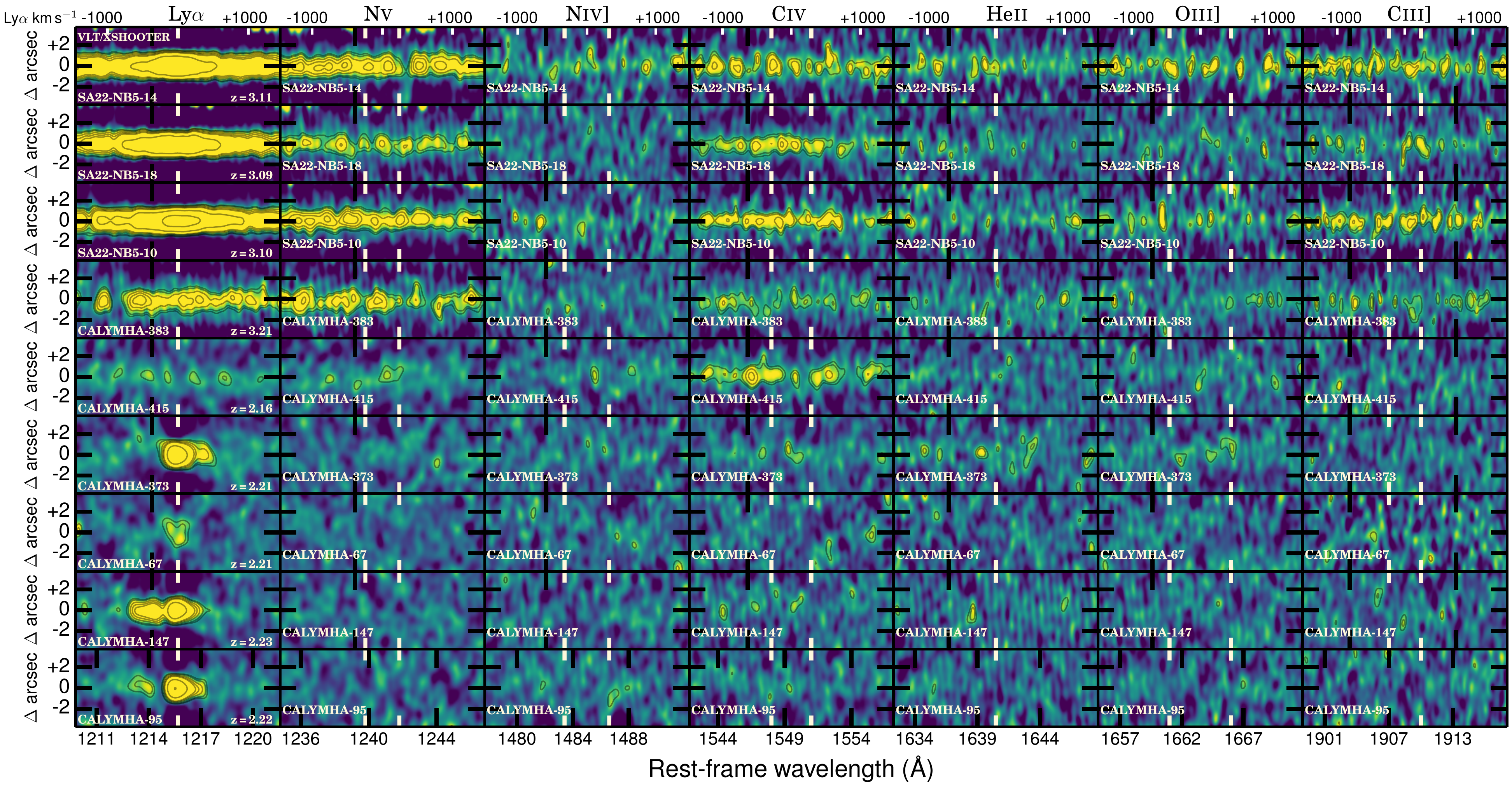}
\caption{Reduced 2D spectra from our VLT/X-SHOOTER observations and zoom in into the major emission lines studied here. Sources are ordered by decreasing Ly$\alpha$ luminosity (top to bottom). We use low and high cut-offs corresponding to $-1$\,$\sigma$ and $+2$\,$\sigma$. We show contours corresponding to $1.5,2,3,4,5,10,20$\,$\sigma$ in all 2D zoom-ins. Velocity offsets are given in respect to Ly$\alpha$. Note that there are significant negative regions (in both flux and signal-to-noise) directly above and below the continuum and emission line detections: this is a simple consequence of jittering along the slit, and a further check for the reliability of the detections (real detections require the negatives up and down). We find that the most luminous LAEs are dominated by broad line AGN, with strong high ionisation rest-frame UV lines, while such lines are much weaker or not present in the faintest among our sample, which show much narrower Ly$\alpha$ profiles.}
\label{2D_full_X-SHOOTER}
\end{figure*}

\subsection{Reduced 2D and 1D spectra}\label{reduced_2D_1D}

We release the binned 1D spectra and we show in figures \ref{2D_full_ISIS}, \ref{2D_full_X-SHOOTER} and \ref{2D_full_DEIMOS} all the 2D spectra for our LAEs, zoomed in at the location of the major emission lines studied in this paper. All velocity offsets shown on the top of each figure are in respect to Ly$\alpha$ (here we use the redshift determined by fitting a Gaussian to the Ly$\alpha$ line). Note that for some lines and for some instrument configurations there is no coverage.

\section{Measuring lines, FWHMs and velocity offsets}\label{FWHMs_and_Voffsets}

By fitting Gaussian line profiles, we use the width of the line ($\sigma_{\lambda_0}$), fitted in \AA \ (rest-frame), and convert it to km\,s$^{-1}$ using:
\begin{equation}
{\rm \sigma}=c\,\frac{\sigma_{\lambda_0}}{\lambda_0} \ \ {\rm km\,s^{-1}}
\end{equation}
where $c$ is the speed of light, 299792.458\,km\,s$^{-1}$ and $\lambda_0$ is the rest-frame wavelength: $\lambda_0=\lambda/(1+z)$. The observed full width at half maximum, FW$_{\rm obs}$, is:
\begin{equation}
{\rm FW_{\rm obs,0}}=2\sqrt{2\times\ln(2)}\times\sigma  \ \ {\rm km\,s^{-1}}.  
\end{equation}

We correct our FW$_{\rm obs}$ measurements for the appropriate instrument dispersion/resolution\footnote{We estimate the instrument resolution by measuring non-blended arc or OH lines.}, FW$_{\rm inst}$ (for Ly$\alpha$: $100-200$\,km\,s$^{-1}$) using:
\begin{equation}
{\rm FWHM}=\sqrt{FW_{\rm obs,0}^2-FW_{\rm inst}^2} \ \ {\rm km\,s^{-1}}. 
\end{equation}
For the Ly$\alpha$ line these lead to corrections of $<100$\,km\,s$^{-1}$ from observed to intrinsic FWHM due to the instrument dispersion. Our results are presented in Table \ref{zspec_Lya_props} and in Section \ref{FWHMs_vs_UV_Lya}.

We use the best fit central wavelength of each line which is significantly detected to obtain a redshift per line ($z_{\rm line}$) and to compute a velocity offset with respect to Ly$\alpha$ by using:

\begin{equation}
{\rm v_{offset,obs}}=c\,\Big(\frac{1+z_{\rm Ly\alpha}}{1+z_{\rm line}}-1\Big) \, \, {\rm km\,s^{-1}}
\end{equation}

We provide tables with the measurements presented and used in this work, e.g. Table \ref{emission_lines_indiv} in a fits format with the final refereed paper, and also the binned spectra.


%
%
%
\begin{table*}
 \centering
  \caption{Measurements and constraints on rest-frame UV emission line fluxes for the sources in our final spectroscopic sample of luminous LAEs. All emission line fluxes are given in $10^{-17}$\,erg\,s$^{-1}$\,cm$^{-2}$. Measurements and the upper and lower errors are derived from perturbing spectra 10,000 times on each individual spectral element and (re-)fitting Gaussian/double Gaussians and obtaining the median, 84th and 16th percentiles, respectively. For sources in which the 0.6 percentile of all 10,000 realisations results in a zero or negative flux we assign the 99.4 percentile as the upper limit, roughly corresponding to a 2.5\,$\sigma$ limit. Fluxes do not include any slit correction. For some sources we either do not have coverage for a specific line and/or the 99.4 percentile flux limit is above $\approx10^{-15}$\,erg\,s$^{-1}$\,cm$^{-2}$ and thus we label those as ``---'' due to being completely unconstrained. ${\rm v_{offset,obs}}$ are measured using detected lines in relation to Ly$\alpha$, including those detected in the IR arm in XSHOOTER data for CALYMHA-147, 373 and 95 (based on [O{\sc iii}] and H$\alpha$; Matthee et al. in prep.)}
\begin{tabular}{@{}cccccccccccc@{}} 
\hline
 ID: Line fluxes & Ly$\alpha$  & N{\sc v}  & N{\sc iv}  & C{\sc iv}   & He{\sc ii}  & O{\sc iii}] & N{\sc iii}] & C{\sc iii}] & ${\rm v_{offset,obs}}$ \\ 
   & (erg\,s$^{-1}$\,cm$^{-2}$)  & & & & & & & & (km\,s$^{-1}$)  \\
\hline
SA22-NB5-14 & $92.0^{+4.0}_{-4.1}$ &  --- &  $<14$ &  $39.5^{+7.0}_{-6.7}$ &  $<5.7$ &  $<13$ &  $<11$ &  $<42$ &  $640^{+340}_{-340}$ & \\
SA22-NB5-18 & $66.4^{+1.9}_{-1.8}$ &  $21.3^{+4.5}_{-3.7}$ &  $<8.6$ &  $44.0^{+9.7}_{-8.2}$ &  $<28$ &  $<9.4$ &  $<25$ &  $8.8^{+3.1}_{-3.8}$ &  $780^{+380}_{-320}$ & \\
SA22-NB5-10 & $86.3^{+2.7}_{-2.7}$ &  --- &  $<7.6$ &  $35.9^{+6.8}_{-6.3}$ &  $<20$ &  $<16$ &  $<36$ &  $17.0^{+4.0}_{-3.8}$ &  $-90^{+230}_{-240}$ & \\
BH-NB5-4(BR3) & $71.4^{+5.1}_{-5.0}$ &  $<55$ & $1.1^{+0.4}_{-0.2}$ &  $14.3^{+0.5}_{-0.5}$ &  $11.1^{+0.2}_{-0.2}$ &  $1.5^{+0.6}_{-0.2}$ &  $1.2^{+0.2}_{-0.2}$ &  $24.0^{+3.6}_{-13.6}$  &  $-20^{+120}_{-130}$ & \\
CALYMHA-383 & $55.2^{+6.3}_{-6.5}$ &  $<80$ &  $<21$ &  $<67$ &  $<15$ &  $<14$ &  $<29$ &  $<46$ &  --- & \\
BH-NB392-12 & $83.6^{+10.5}_{-10.0}$ &  $<32$ &  --- &  --- &  --- &  --- &  --- &  --- &  --- & \\
BH-NB5-6 & $13.2^{+1.2}_{-1.1}$ &  $6.2^{+1.0}_{-1.1}$ &  --- &  $<14$ &  $<12$ &  $<10$ &  $<9.9$ &  $25.1^{+6.0}_{-5.8}$ &  $-90^{+400}_{-400}$ & \\
BH-NB5-10 & $12.7^{+0.2}_{-0.2}$ &  $2.1^{+3.2}_{-0.3}$ &  $<8.2$ &  $<1.2$ &  $2.6^{+0.3}_{-0.3}$ &  --- &  $<0.7$ &  $<5.3$ &  $150^{+170}_{-500}$ & \\
BH-NB5-16 & $10.6^{+0.3}_{-0.3}$ &  $6.6^{+0.7}_{-0.7}$ &  $2.6^{+0.6}_{-1.5}$ &  $<9.0$ &  $<0.6$ &  $<1.3$ &  $2.0^{+9.1}_{-1.2}$ &  $<9.1$ &  $110^{+580}_{-180}$ & \\
GN-NB5-6712 & $26.6^{+4.7}_{-4.0}$ &  $<65$ &  --- &  $9.2^{+2.1}_{-2.3}$ &  $6.6^{+2.9}_{-2.3}$ &  $<23$ &  $<17$ &  --- &  $20^{+160}_{-110}$ & \\
BH-NB392-55 & --- &  --- &  $<3.2$ &  $34.7^{+99.0}_{-8.6}$ &  $<14$ &  $<15$ &  --- &  --- &  --- & \\
BH-NB5-27 & $13.1^{+0.3}_{-0.3}$ &  $1.3^{+0.4}_{-0.3}$ &  $<13$ &  $<0.7$ &  $<4.1$ &  $<1.9$ &  $<2.8$ &  $<1.3$ &  $-160^{+230}_{-170}$ & \\
BH-NB5-37 & $10.7^{+0.2}_{-0.2}$ &  $<1.8$ &  $<22$ &  $<3.8$ &  $0.7^{+0.3}_{-0.2}$ &  $<3.7$ &  $<0.5$ &  $<3.2$ &  $-110^{+90}_{-100}$ & \\
BH-NB5-34 & $4.1^{+0.4}_{-0.4}$ &  $<8.5$ &  $<0.5$ &  $8.2^{+0.6}_{-0.6}$ &  $<3.3$ &  $<0.4$ &  $<1.7$ &  --- &  $460^{+170}_{-190}$ & \\
CALYMHA-147 & $43.0^{+2.9}_{-3.7}$ &  $<18$ &  $<12$ &  $<7.8$ &  $<15$ &  $<23$ &  $<37$ &  --- &  $250^{+10}_{-10}$ & \\
CALYMHA-373 & $12.8^{+0.7}_{-0.7}$ &  $<4.4$ &  $<2.6$ &  $<2.0$ &  $<2.2$ &  $<5.0$ &  $<55$ &  $<12$ &  $150^{+10}_{-10}$ & \\
CALYMHA-415 & $25.5^{+6.9}_{-7.2}$ &  $12.7^{+4.5}_{-4.0}$ &  $<17$ &  $15.8^{+3.6}_{-3.2}$ &  $<6.3$ &  $<5.2$ &  --- &  $<59$ &  $-20^{+350}_{-350}$ & \\
CALYMHA-67 & $8.4^{+1.9}_{-1.7}$ &  $<11$ &  $<15$ &  $<11$ &  $<5.7$ &  $<5.2$ &  $<39$ &  $<56$ &  --- & \\
GN-NB5-3378 & $2.7^{+0.1}_{-0.1}$ &  $<0.6$ &  $<1.0$ &  $<1.1$ &  $0.6^{+0.3}_{-0.2}$ &  $<0.3$ &  $<1.9$ &  $<1.3$ &  $-250^{+760}_{-650}$ & \\
CALYMHA-95 & $15.6^{+0.8}_{-0.7}$ &  $<11$ &  $<7.6$ &  $<7.6$ &  $<1.7$ &  $<6.0$ &  $<33$ &  $<10$ &  $30^{+70}_{-60}$ & \\
GN-NB5-5878 & $4.7^{+0.2}_{-0.2}$ &  $<0.5$ &  $<0.8$ &  $<0.2$ &  $<0.3$ &  $<0.5$ &  $<1.0$ &  $<1.8$ &  --- & \\

\hline
\#Lines detected  &  20 (100\%) & 6 (33\%)  & 2 (11\%) & 8 (40\%) & 5 (25\%) & 1 (5\%) & 2 (11\%) & 4 (25\%)  & --- \\
\hline
\end{tabular}
\label{emission_lines_indiv}
\end{table*}

\section{{\sc cloudy} modelling}\label{CLOUDY_AP}

We explore stellar-like ionising spectra by using a large range of black bodies, with effective temperatures extending from 20,000 K to 160,000 K in steps of 1,000\,K. We explore a wide range of metallicities, from 0.01\,Z$_{\odot}$ to 3\,Z$_{\odot}$ in 0.05 steps in log space. We also vary the ionisation parameter $U$, from 0.00001 to 10 ($-5$ to $2$ in log in steps of 0.2), from lower to high ionisation parameters. We follow known relations between how C/O varies with metallicity, but also run {\sc cloudy} for different C/O ratios allowed by current data \citep[see][]{Nakajima2017}. We use densities of 100, 300 and 1000\,cm$^{-2}$. Here we focus on the emerging emission line ratios. We note that our aim is not to claim ``true" physical conditions, but rather to broadly identify which physical conditions may be present. In our derived best physical conditions we also take all the models within 1\,$\sigma$ of the observed line ratios, or that are allowed by the line ratio limits.

In order to compare with more realistic stellar models, we use {\sc bpass} \citep[][]{Eldridge2009,Eldridge2012,Stanway2016}\footnote{See also \cite{2017PASA...34...58E}.} models that include stellar rotation and binary evolution. These allow us to better constrain realistic ionising fluxes from stellar populations and to provide a link to indicative stellar ages. Furthermore, we also explore power-law ionising sources, which we associate with AGN, and run those models for the same physical conditions. For many emission line ratios the hottest black body ionising sources become very similar to the result of a power-law ionising source, but for the even higher ionisation lines, and for appropriate choices of lines, there is still a good separation. We note nonetheless, that naturally some of the line ratios used to separate stellar from an AGN nature that have used standard stellar populations do not work in the cases where stars can have much higher effective temperatures (due to binary interaction, winds exposing the deeper components of the star, or rotation).

\bsp	
\label{lastpage}
\end{document}